\documentclass[12pt]{article}
\usepackage{amsmath, amssymb}
\usepackage{graphicx,psfrag,epsf}
\usepackage{enumerate}
\usepackage{natbib}

\usepackage{url} 

\usepackage[bookmarks=true, colorlinks
=true, linkcolor = blue, citecolor=blue]{hyperref}

\newcommand{\blind}{1}

\usepackage[hmargin=1in,vmargin=1in]{geometry}

\allowdisplaybreaks 

\newtheorem{assumption}{Assumption}
\newtheorem{lemma}{Lemma}
\newtheorem{theorem}{Theorem}

\usepackage{longtable}  
\usepackage{siunitx}    
\usepackage{booktabs}
\usepackage{ragged2e}
\begin{document}

\def\spacingset#1{\renewcommand{\baselinestretch}%
{#1}\small\normalsize} \spacingset{1}


\if1\blind
{
  \title{\bf Difference-in-Differences with Time-varying Continuous Treatments Using Double/Debiased Machine Learning\thanks{
    We gratefully acknowledge comments from Michael Knaus and participants at the Valais Econometrics Workshop 2025. 
    Addresses for correspondence: Michel F. C. Haddad, Bancroft Building, Queen Mary University of London, Mile End Rd, London E1 4NS, United Kingdom; m.haddad@qmul.ac.uk. Martin Huber, University of Fribourg, Bd de P\'erolles 90, 1700 Fribourg, Switzerland; martin.huber@unifr.ch. Lucas Z. Zhang, Bates White, 2001 K Street NW
North Building, Suite 500, Washington, DC 20006, USA; lucaszz@g.ucla.edu.
    }}
  \author{Michel F. C. Haddad\\
    \small{Queen Mary University of London, Dept. of Business Analytics and Applied Economics} \vspace{0.2 cm}\\
    Martin Huber\\
    \small{University of Fribourg, Dept.\ of Economics} \vspace{0.2 cm}\\
    José Eduardo Medina-Reyes\\
    \small{Queen Mary University of London, Dept. of Business Analytics and Applied Economics} \vspace{0.2 cm}\\
    Lucas Z. Zhang\\
    \small{University of California, Los Angeles, Dept.\ of Economics; Bates White Economic Consulting}}
     \date{}
  \maketitle
} \fi

\begin{abstract}
We propose a difference-in-differences (DiD) framework designed for time-varying continuous treatments across multiple periods. Specifically, we estimate the average treatment effect on the treated (ATET) by comparing distinct non-zero treatment intensities. Identification rests on a conditional parallel trends assumption that accounts for observed covariates and past treatment histories. Our approach allows for lagged treatment effects and, in repeated cross-sectional settings, accommodates compositional changes in covariates. We develop kernel-based ATET estimators for both repeated cross-sections and panel data, leveraging the double/debiased machine learning framework to handle potentially high-dimensional covariates and histories. We establish the asymptotic properties of our estimators under mild regularity conditions and demonstrate via simulations that their undersmoothed versions perform well in finite samples. As an empirical illustration, we apply our estimator to assess the effect of the second-dose COVID-19 vaccination rate in Brazil and find that higher vaccination rates reduce COVID-19-related mortality after a lag of several weeks.

\end{abstract}

\noindent%
{\it Keywords:} Causal inference, Continuous treatment, Difference-in-differences, Machine learning, Parallel trends.
\vfill

	\thispagestyle{empty}
	\newpage
	\setcounter{page}{1}

\spacingset{1.45} 
\section{Introduction}
\label{sec:intro}

Difference-in-differences (DiD) is a cornerstone method for causal inference, i.e., the evaluation of the impact of a treatment (such as vaccination) on an outcome of interest (such as mortality), in observational studies, provided that outcomes may be observed both before and after the treatment introduction. In such studies, the observed outcomes of treated individuals before and after a treatment typically do not allow researchers to directly infer the treatment effect. This is due to confounding time trends in the treated individuals' counterfactual outcomes that would have occurred without the treatment. The DiD approach tackles this issue based on a control group not receiving the treatment and the so-called parallel trends assumption, imposing that the treated group's unobserved counterfactual outcomes under non-treatment follow the same observed mean outcome trend of the control group. While the canonical DiD setup considers a binary treatment definition (treatment versus no treatment), in many empirical applications, treatments are continuously distributed - e.g., the vaccination rate in a region. Furthermore, the treatment intensity might vary over time and a strict control group with a zero treatment might not be available across all or even any treatment period, as also argued in \cite{dechaisemartin2023differenceindifferences}, who point to taxes, tariffs, or prices as possible treatment variables with strictly non-zero doses.

To address these limitations, we propose an DiD approach for a time-varying continuous treatment across multiple time periods, enabling the assessment of the average treatment effect on the treated (ATET) when comparing two non-zero treatment doses, such as a higher versus a lower vaccination rate in a region. Identification relies on a conditional parallel trend assumption imposed on the mean potential outcome under the lower dose, given observed covariates and past treatment histories. This assumption requires that, conditional on covariates and previous treatments, the counterfactual outcomes of the treatment group (receiving the higher treatment dose) under the lower dose follow the same mean outcome trend as in the control group (receiving the lower treatment dose). We note that this parallel trend assumption on counterfactual outcomes under a positive treatment dose is more stringent than that on non-treated outcomes, see e.g., the discussion in \cite{Fricke2017}, which is the trade-off for DiD-based identification of causal effects based on comparisons of non-zero treatment doses.

We suggest novel semiparametric kernel-based DiD estimators for both repeated cross-sections, in which subjects differ across time periods, and panel data, in which the same subjects are repeatedly observed across periods. The estimators adopt the double/debiased machine learning (DML) framework introduced in \cite{Chetal2018} to control for covariates and past treatment histories in a data-adaptive manner, which appears attractive in applications where the number of potential control variables is large relative to the sample size. Thus, we estimate the models for the conditional mean outcomes and conditional treatment densities - also known as generalized treatment propensity scores, based on machine learning and use these nuisance parameters as plug-ins in doubly robust (DR) score functions (see for instance \cite{Robins+94} and \cite{RoRo95}, which are tailored to DiD-based ATET estimation).

These DR scores include kernel functions on the continuous treatment as also considered in \cite{Kennedyetal2017} and \cite{colangelo2023doubledebiasedmachinelearning}, when imposing a selection-on-observables-assumption (rather than a parallel trend assumption) for identification. Under an appropriately shrinking bandwidth as the sample size increases, we demonstrate that our estimators satisfy \citeauthor{Neyman1959}-orthogonality, implying that they are substantially robust to approximation errors in the nuisance parameters under specific regularity conditions - e.g., approximate sparsity when using lasso regression \citep{Tibshirani96} for nuisance parameter estimation as considered in \cite{Bellonietal2014}). Following \cite{Chetal2018}, we also apply cross-fitting such that the nuisance parameters and scores are not simultaneously estimated in the same parts of the data, to avoid overfitting. To account for the biases introduced by the kernel functions, we use undersmoothed bandwidths, implying that the biases vanish asymptotically. We show that under particular regularity conditions, the DML kernel estimators satisfy asymptotic normality, which also provides us with an asymptotic formula for variance estimation. 

We also conduct a simulation study, which points to a compelling finite sample behavior of undersmoothed versions of our estimators under a sample size of several thousand observations. As an empirical application, we use Brazilian municipality-level panel data spanning March 2020 to May 2022 to estimate the effect of the continuously distributed second-dose COVID-19 vaccination rate on mortality. Comparing vaccination rates of 60\% versus 40\%, we find that higher vaccination coverage does not have a statistically significant effect on COVID-19-related mortality in the first two weeks. However, it leads to a statistically significant reduction in mortality after approximately four weeks, consistent with the expected time lag between infection and death.


Our study adds to a growing literature on the semi- and nonparametric DiD-based identification of causal effects of continuously distributed treatments. This framework avoids potential misspecification errors of classical linear two-way fixed effects models - see, for instance, the discussion in \cite{de2020two}. \cite{callaway2024difference} consider a treatment that has a mass at zero (constituting control group), but might also take continuously distributed non-zero doses. This permits identifying the ATET of a specific non-zero treatment dose versus no treatment, under the parallel trend assumption that the treatment group (receiving a non-zero dose) and the control group share the same trend in mean outcomes under non-treatment. However, \cite{callaway2024difference} do not consider semi- or nonparametric estimation conditional on covariates. \cite{zhang2025} adopts a conditional parallel trends assumption in the continuous treatment context, implying that parallel trends are assumed to hold conditional on observed covariates, as previously introduced by \cite{abadie2005} for binary treatments. \cite{zhang2025} studies the canonical two-period model (with one pre- and one post-treatment period) in both the panel and repeated cross-section data, providing identification, estimation, and inference results using the double/debiased machine learning (DML) framework. In a related work, \cite{Hettingeretal2025} combine two parallel trends assumptions to identify the average dose effect on the entire treated group in the panel setting. They adopt a similar framework as \cite{Kennedyetal2017} to establish a multiply-robust estimator.

Our paper extends these results in several ways. First, we permit comparisons between two non-zero treatment doses under our stronger parallel trends assumption. Second, we allow for more than two time periods with possibly non-zero treatment doses that may vary over time, such as in a staggered adoption scenario - as e.g.,\ considered for binary treatments in \cite{GoodmanBacon2018}, \cite{SantAnnaZhao2018}, \cite{sun2021estimating}, and \cite{borusyak2024revisiting}), which motivates conditioning on past treatment histories when assessing the ATET. Third, our approach for repeated cross-sections is capable of appropriately tackling covariate distributions that change within treatment groups over time, while many established approaches require them to be time-invariant, as pointed out in \cite{Hong2013}, which concerns e.g.,\ the DID methods for binary or discrete treatments in \cite{abadie2005} using inverse probability weighting, \cite{SantAnnaZhao2018} using DR, or \cite{Chang2020} using DML. A notable exception is the DML approach in \cite{Zimmert2018}, which allows for time-varying covariate distributions within treatment groups, nonetheless applying to binary rather than continuous treatments.

We also distinguish our results from \cite{dechaisemartin2023differenceindifferences}, which considers treatments with strictly non-zero doses, and discuss effect identification and estimation under parallel trend assumptions related to those considered in the present paper. The authors focus on evaluating an average ATET across the treatment margins of all subjects of changing their treatment dose, which are considered as treatment dose - e.g., what is the average effect of marginally increasing the vaccination rate on mortality among all regions that increase the vaccination rate by possibly different margins. In contrast, our study suggests estimators for assessing the ATET based on comparing two specific treatment doses rather than averaging across many doses - e.g., the effect of a vaccination rate of 40\% versus 30\%. As a further distinction, \cite{dechaisemartin2023differenceindifferences} do not consider controlling for covariates when imposing parallel trends, while our estimators permit controlling for covariates in a data-driven manner through machine learning.

The remainder of the paper is organized as follows. Section \ref{sec:id} motivates and establishes the identification of the instantaneous ATET for continuous treatments in repeated cross-sections. Section \ref{sec:id2} generalizes this framework to time-varying treatments, accommodating lagged effects and dynamic responses. Section \ref{paneldata} addresses identification in panel data settings. Section \ref{sec:meth} introduces our proposed estimators based on double machine learning (DML) with cross-fitting, while Section \ref{asymp} derives their asymptotic properties. Section \ref{sim} assesses finite-sample performance through a simulation study. Section \ref{appl} presents an empirical application evaluating the effectiveness of the second-dose COVID-19 vaccination rate in Brazil on mortality. Section \ref{sec:conc} concludes.

\section{Instantaneous effects in repeated cross-sections}\label{sec:id}

This section discusses the identification of the instantaneous ATET, evaluated shortly after treatment introduction, in the repeated cross-sectional setting. For intuition, we first consider a discrete treatment, denoted by $D$. We first assume that there is a single point in time at which the treatment switches from zero to a non-zero treatment dose for a subset of the population (i.e., the treated subjects), while it remains at zero for another subset which constitutes the control group. Accordingly, let $T$ denote the time period, where $T=0$ is the baseline period prior to treatment assignment, while $T>0$ is a post-treatment period after the introduction of a positive treatment dose among the treated. The treatment dose may differ across subjects and for now it is assumed to remain constant across post-treatment periods. Let $Y_t$ denote the observed outcome at period $T=t$, while $Y_t(d)$ is the potential outcome at period $t$ when hypothetically setting the treatment to $D=d$, see e.g.,\ \cite{Neyman23} and \cite{Rubin74} for further discussion on the potential outcome framework. Throughout the present work, we impose the stable unit treatment value assumption (SUTVA), meaning that the potential outcomes of a subject are not affected by the treatment of others - see, for instance, the discussion in \cite{Rubin80} and \cite{Cox58}. This permits defining the average treatment effect on the treated (ATET) of treatment dose $d$ in period $t$ as \begin{align}\label{ATEsimple}\Delta_{d,t}=E[Y_t(d)-Y_t(0)|D=d, T=t].\end{align} 
\noindent Furthermore, let $X$ denote the observed covariates. We impose the following identifying assumptions, which are closely related to those in \cite{Lechner2010}.

\begin{assumption} {\bf (Conditional parallel trends):}\label{ass1}\\
$E[Y_t(0)-Y_0(0)|D=d,T=t,X]=E[Y_t(0)-Y_0(0)|D=0,T=t,X]$ for $d$ in the support of $D$ and $t>0$ in the support of $T$.
\end{assumption}

\begin{assumption} {\bf (No anticipation):}\label{ass2}\\
$E[Y_0(d)-Y_0(0)|D=d,T=0,X]=0$ for $d > 0$ in the support of $D$.
\end{assumption}

\begin{assumption}{\bf (Common support):}\label{ass3}\\
$\Pr(D = d,\, T = t \mid X,\,(D, T) \in \{(d', t'), (d, t)\}) < 1
\quad \text{for } d > 0 \text{ in the support of } D,\;
t > 0 \text{ in the support of } T,\;
\text{and } (d', t') \in \{(d, 0), (0, t), (0, 0)\}.$
\end{assumption}

\begin{assumption} {\bf (Exogenous covariates):}\label{ass4}\\
$X(d)=X$ for all $d$ in the support of $D$.
\end{assumption}

Assumption \ref{ass1} imposes parallel trends in the mean potential outcome under non-treatment across treated and control groups, conditional on covariates $X$. Assumption \ref{ass2} imposes that conditional on $X$, the ATET must be zero in the pre-treatment period $T=0$, which on average rules out non-zero anticipation effects among the treated in period $T=0$.  Assumption \ref{ass3}  requires that for any subject receiving treatment dose $d$ in a post-treatment period $t > 0$, there exist in terms of $X$ comparable treated subjects in the baseline period, non-treated subjects in the post-treatment period, and non-treated subjects in the baseline period. Lastly, Assumption \ref{ass4} imposes that the covariates are not causally affected by the treatment. We note that this may be a relevant restriction if the researchers include post-treatment variables - as it is frequently the case in repeated cross-sections (see, for instance, the discussion in \cite{caetano2022difference}. 

Under these assumptions, the conditional ATET of treatment dose $d>0$ in period $t>0$ given $X$ is identified:
\begin{eqnarray}\label{DiDobs}
&&E[Y_t(d)-Y_t(0)|D=d,T=t,X]\\ 
&=&E[Y_t(d)|D=d,T=t,X]-E[Y_0(0)|D=d,T=0,X]\notag\\
&-&E[Y_t(0)|D=d,T=t,X]+E[Y_0(0)|D=d,T=0,X]\notag\\
&=&E[Y_t(d)|D=d,T=t,X]-E[Y_0(d)|D=d,T=0,X]\notag\\
&-&\{E[Y_t(0)|D=d,T=t,X]-E[Y_0(0)|D=d,T=0,X]\}\notag\\
&=&E[Y_t(d)|D=d,T=t,X]-E[Y_0(d)|D=d,T=0,X]\notag\\
&-&\{E[Y_t(0)|D=0,T=t,X]-E[Y_0(0)|D=0,T=0,X]\}\notag\\
&=&E[Y_T|D=d,T=t,X]-E[Y_T|D=d,T=0, X]\notag\\
&-&\{E[Y_T|D=0,T=t,X]-E[Y_T|D=0, T=0, X]\}.\notag
\end{eqnarray}
The first equality in \eqref{DiDobs} follows from subtracting and adding $E[Y_0(0)|D=d,X]$, and the second from Assumption \ref{ass2}, implying that $E[Y_0(0)|D=d,T-0,X]=E[Y_0(d)|D=d,T=0,X]$. The third equality follows from the conditional parallel trends assumption (Assumption \ref{ass1}). The fourth equality follows from the fact that $Y_T=Y_t(d)$ conditional on $D=d$ and $T=t$, which is known as the observational rule. By Assumption \ref{ass4}, the effect of the treatment on the outcome $Y_t$ is not mediated by $X$, such that the conditional ATET in expression \eqref{DiDobs} captures the total causal effect among the treated given $X$, rather than only a partial effect not mediated by $X$. Averaging the conditional ATET over the distribution of $X$ among the treated in a post-treatment period $T=t$ yields the ATET in that period:
\begin{eqnarray}\label{DiDidentpostreg}
&&\Delta_{d,t}=E[  \mu_d(t,X)-\mu_d(0,X) - (\mu_0(t,X)-\mu_0(0,X))|D=d, T=t ],
\end{eqnarray}
where we use the short hand notation $\mu_d(t,x)=E[Y_T|D=d,T=t,X=x]$ for the conditional mean outcome given the treatment, the time period, and the covariates. 

If the treatment $d$ consisted of discrete treatment doses, the ATET can alternatively be assessed based on the following doubly robust (DR) expression, which is in analogy to \cite*{Zimmert2018} for binary treatments:
\begin{align}\label{DiDidentDR}
&\Delta_{d,t}=E\left[  \frac{I\{D=d\}\cdot I\{T=t\}\cdot [Y_T-\mu_d(0,X)-\mu_0(t,X)+\mu_0(0,X) ] }{ \Pi_{d,t}} \right.\notag\\
&- \left\{\frac{I\{D=d\}\cdot I\{T=0\}\cdot \rho_{d,t}(X)}{ \rho_{d,0}(X)\cdot \Pi_{d,t}}\right.\\
&- \left.\left.\left(\frac{I\{D=0\}\cdot  I\{T=t\}\cdot \rho_{d,t}(X)}{ \rho_{0,t}(X) \cdot \Pi_{d,t}} - \frac{I\{D=0\}\cdot I\{T=0\}\cdot \rho_{d,t}(X)}{ \rho_{0,0}(X)\cdot \Pi_{d,t}}\right) \right\} \cdot (Y_T-\mu_D(T,X))  \right],\notag
\end{align}
where $\Pi=\Pr(D=d,T=t)$ denotes the unconditional probability of receiving treatment dose $d$ at period $t>0$, $\rho_{d,t}(X)=\Pr(D=d,T=t|X)$ denotes the conditional probability of a specific treatment group-period-combination $D=d,T=t$ given $X$, and $I\{\cdot\}$ denotes the indicator function.

To see that the DR expression \eqref{DiDidentDR} indeed corresponds to the ATET in equation \eqref{DiDidentpostreg},
note that the first term on the right of equation \eqref{DiDidentDR}, 
by the law of iterated expectation, corresponds to the ATET:
\begin{align*}
&E\left[  \frac{I\{D=d\}\cdot I\{T=t\}\cdot [Y_T-\mu_d(0,X)-\mu_0(t,X)+\mu_0(0,X) ] }{ \Pi_{d,t}} \right]\\
&= E[  Y_T-\mu_d(0,X) - (\mu_0(t,X)-\mu_0(0,X))|D=d, T=t ]\\
&= E[  \mu_d(t,X)-\mu_d(0,X) - (\mu_0(t,X)-\mu_0(0,X))|D=d, T=t ].
\end{align*}
Moreover, again by the law of iterated expectation, we have for any $d'$ and $t'$ in the support of $D$ and $T$,
\begin{align*}
&E\left[ \frac{I\{D=d'\}\cdot I\{T=t'\} \cdot \rho_{d,t}(X)} {\rho_{d',t'}(X) \cdot \Pi_{d,t}}  \cdot (Y_T-\mu_D(T,X))\right]\notag\\
&=E\left[ \frac{\rho_{d,t}(X)}{\Pi_{d,t}} \cdot E\left[ \frac{I\{D=d'\}\cdot I\{T=t'\}\cdot (Y_T-\mu_D(T,X))}{\rho_{d',t'}(X)}\Big| X\right]  \right]\notag\\
&=E\left[ \frac{\rho_{d,t}(X)}{\Pi_{d,t}} \cdot E [ Y_T-\mu_D(T,X) | D=d', T=t', X]  \right]\notag\\
&= E [\mu_{d'}(t',X)-\mu_{d'}(t',X) | D=d,T=t ]=0.
\end{align*}
Therefore, the remaining terms in equation \eqref{DiDidentDR} are all equal to zero. This demonstrates that \eqref{DiDidentpostreg} and \eqref{DiDidentDR} are equivalent and both yield the ATET under our identifying assumptions.  

To adapt the previous results based on discrete treatments into a continuous treatment, the indicator functions for non-zero treatment doses $d$ need to be replaced by smooth weighting functions with a bandwidth approaching zero to obtain identification. 
Thus, we denote by $\omega(D;d,h)$ a weighting function that depends on the distance between $D$ and the non-zero reference value $d$ as well as a non-negative bandwidth parameter $h$: $\omega(D;d,h)=\frac{1}{h}\mathcal{K}\left(\frac{D-d}{h}\right)$, with $\mathcal{K}$ denoting a well-behaved kernel function. The closer $h$ is to zero, the less weight is given to realizations of $D$ that are further away from $d$. 
Moreover, we note that the previous (conditional) probabilities $\Pi_{d,t}$ and $\rho_{d,t}(X)$ for $d>0$ and $t\geq0$ should be replaced by (conditional) density functions if the treatment is continuous. Note that the conditional treatment density is also referred to as generalized propensity score in the literature, e.g., \citet{ImaivanDyk2004} and \citet{HiranoImbens2005}. In fact, as observed in \citet{fan1996estimation} the conditional densities can also be expressed as the limit of the conditional expectations of kernel functions: $\rho_{d,t}(X)=\lim_{h \rightarrow 0} E[\omega(D;d,h) \cdot I\{T=t\}|X]$. Similarly, $\Pi_{d,t}=\lim_{h \rightarrow 0} E[\omega(D;d,h) \cdot I\{T=t\}]$. 
Applying these modifications to DR expression \eqref{DiDidentDR} yields
\begin{align}\label{DiDidentDRcont}
&\Delta_{d,t}= \lim_{h \rightarrow 0} E\left[  \frac{  \omega(D;d, h) \cdot I\{T=t\}\cdot [Y_T-\mu_d(0,X)-\mu_0(t,X)+\mu_0(0,X)]}{ \Pi_{d,t}}\right.\notag\\
&- \frac{\omega(D;d, h) \cdot I\{T=0\}\cdot \rho_{d,t}(X)\cdot \epsilon_{d,0,X}}{ \rho_{d,0}(X)\cdot \Pi_{d,t}}\notag\\
&- \left.\left(\frac{I\{D=0\}\cdot  I\{T=t\}\cdot \rho_{d,t}(X)\cdot \epsilon_{0,t,X}}{ \rho_{0,t}(X) \cdot \Pi_{d,t}} - \frac{I\{D=0\}\cdot I\{T=0\}\cdot \rho_{d,t}(X)\cdot \epsilon_{0,0,X}}{ \rho_{0,0}(X)\cdot \Pi_{d,t}}\right) \right],
\end{align}
where $\epsilon_{d,t,x}=(Y_T-\mu_d(t,x))$ denotes a regression residual. It is worth mentioning that in a related study, \cite{zhang2025} also provides an DR expression for $\Delta_{d,t}$ with continuous treatment doses $d$ that in the repeated cross-sections setting. However, in contrast to our results, his approach is not applicable to settings in which the distribution of covariates $X$ changes over time. 

Section \ref{sec:id2} extends the analysis to the identification of ATETs in later periods following treatment, allowing for time lags and potentially time-varying treatment doses.

\section{Time-varying treatment doses in repeated cross-sections}\label{sec:id2}

In this section, we extend our DiD framework to accommodate time-varying treatment doses and multiple treatment (and, later, outcome) periods. For notation, we denote the treatment dose in period $t$ by $D_t$, and we use $\mathbf{D}_{t}=\{D_0,...,D_t\}$ to denote the sequence of treatment doses up to and including period $t$. We replace Assumption \ref{ass1}, which imposes parallel trends on the mean potential outcome under non-treatment, with Assumption \ref{ass5}, which imposes parallel trends in mean potential outcomes between non-zero treatment doses.
\begin{assumption} {\bf (Conditional parallel trends with non-zero treatment doses):}\label{ass5}\\
$E[Y_t(d_t')-Y_{t-1}(d_{t-1}'')|D_t=d_t,T=t,\mathbf{D}_{t-1}=d_{t-1}'',X] \\ = E[Y_t(d_t')-Y_{t-1}(d_{t-1}'')|D_t=d_t',T=t,\mathbf{D}_{t-1}=d_{t-1}'',X]$ for  $d_t>d_t'\geq 0$ in the support of $D_t$,   $\mathbf{d}_{t-1}''$ in the support of $\mathbf{D}_{t-1}$, and $t>0$ in the support of $T$.
\end{assumption}

Assumption \ref{ass5} states that for two groups that receive a higher treatment dose $d_t$ (treatment group) and a lower treatment dose $d_t'$ (control group), respectively, in period $t$, and both receive treatment dose $d_{t-1}''$ in period $t-1$, the trend in mean potential outcomes when moving from dose $d_{t-1}''$ in period $t-1$ to dose $d_t'$ in period $t$ is the same across groups, conditional on $X$. For the special case where $d_t' = d_{t-1}''$, see for example \citet{dechaisemartin2023differenceindifferences} and \citet{dechaisemartin2025treatmenteffectestimationcomplexdesigns}, this implies that if no group changed the treatment dose between periods $t-1$ and $t$, then the treatment and control groups would experience the same trend in their mean outcomes. Moreover, we note that Assumption \ref{ass5} is stronger than Assumption \ref{ass1} because it imposes parallel trends with respect to non-zero treatment doses. 

This generally imposes restrictions on treatment effect heterogeneity; see the discussions in \citet{Fricke2017} and \citet{deChaisemartin2022did}. For instance, if Assumption \ref{ass1} is plausibly satisfied, then Assumption \ref{ass5} will additionally hold only if the change in the average treatment effects of receiving dose $d_t'$ (relative to no treatment, $d_t=0$) in period $t$ and receiving dose $d_{t-1}''$ (relative to $d_{t-1}=0$) in period $t-1$ is identical across the treatment group receiving $d_t$ and the control group receiving $d_t'$ in period $t$, conditional on $\mathbf{D}_{t-1}=d_{t-1}''$ and $X$. Two special cases satisfying this condition are that (i) average treatment effects are homogeneous across groups in each period, or (ii) average treatment effects are time-invariant within groups (though levels may differ across groups).\footnote{To show this formally, consider the conditional expectation in the first line of Assumption \ref{ass5}, $E[Y_t(d_t')-Y_{t-1}(d_{t-1}'') \mid D_t=d_t,T=t,\mathbf{D}_{t-1}=d_{t-1}'',X]$, and add and subtract $Y_t(0)$ as well as $Y_{t-1}(0)$, which yields
\[
\underbrace{E[Y_t(0) - Y_{t-1}(0) \mid D_t=d_t, \mathbf{D}_{t-1}=d_{t-1}'', X]}_{=0\textrm{ under Assumption \ref{ass1}}} + \tau_{d_t,d_{t-1}'', X}^t(d_t') - \tau_{d_t,d_{t-1}'', X}^{t-1}(d_t''),
\]
where $\tau_{d_t,d_{t-1}'', X}^t(d_t') = E[Y_t(d_t') - Y_t(0) \mid D_t=d_t, \mathbf{D}_{t-1}=d_{t-1}'', X]$ denotes the conditional average treatment effect of moving from $0$ to $d_t'$ at time $t$. Conditional on Assumption \ref{ass1}, Assumption \ref{ass5} holds if and only if the change in the average treatment effects over time is identical across treatment groups $d_t$ and $d_t'$:
\[
(\tau_{d_t,d_{t-1}'', X}^t(d_t') - \tau_{d_t,d_{t-1}'', X}^{t-1}(d_t'')) - (\tau_{d_t',d_{t-1}'', X}^t(d_t') - \tau_{d_t',d_{t-1}'', X}^{t-1}(d_t'')) = 0.
\]
This is satisfied in two special cases: (i) effect homogeneity across groups within each period, i.e.\ $\tau_{d_t,d_{t-1}'', X}^t(d_t')=\tau_{d_t',d_{t-1}'', X}^t(d_t')$ and $\tau_{d_t,d_{t-1}'', X}^{t-1}(d_t'')=\tau_{d_t',d_{t-1}'', X}^{t-1}(d_t'')$; or (ii) time-invariant effects within groups, i.e.\ $\tau_{d_t,d_{t-1}'', X}^t(d_t')=\tau_{d_t,d_{t-1}'', X}^{t-1}(d_t'')$ and $\tau_{d_t',d_{t-1}'', X}^t(d_t')= \tau_{d_t',d_{t-1}'', X}^{t-1}(d_t'')$.}

Furthermore, we replace the previous Assumption \ref{ass2} (ruling out anticipation) by the following condition:
\begin{assumption} {\bf (No anticipation):}\label{ass6}\\
$E[Y_{t-1}(d_t)-Y_{t-1}(d_t')|D=d_t,T=t-1,\mathbf{D}_{t-1},X]=0$ for all $d_t\neq d_t'$ in the support of $D_t$.
\end{assumption}
Lastly, the previous common support condition given in Assumption \ref{ass3} is replaced by the following assumption, where $f(A|B)$ denotes the conditional density of $A$ given $B$:  
\begin{assumption}{\bf (Common support):}\label{ass7}\\
$f(D_t = d_t,\, T = t \mid \mathbf{D}_{t-1}, X,\, (D_t, T) \in \{(d^*_t, t'), (d_t, t)\}) < 1
\quad \text{for } d_t > d'_t \text{ in the support of } D_t,\;
t > 0 \text{ in the support of } T,\;
\text{and } (d^*_t, t') \in \{(d_t, t-1), (d'_t, t), (d'_t, t-1)\}.$
\end{assumption}

We note that our modified set of Assumptions \ref{ass4} to \ref{ass7} does not rule out that treatment doses in earlier periods affect outcomes in later periods. Thus, the treatment sequence $\mathbf{D}_{t-1}$ may influence the outcome in period $t$. In this case, we may nevertheless identify the ATET of treatment dose $d_t$ versus $d_t'$ in period $t$ net of any direct effects of the previous history of treatment doses on the outcome in period $t$, formally defined as 
\begin{align}\label{ATEsameperiod}\Delta_{d_t,d_t', t}=E[Y_t(d_t)-Y_t(d_t')|D_t=d_t,T=t].\end{align}
\noindent It is worth noticing that under Assumptions \ref{ass4} to \ref{ass7}, the conditional ATET of treatment dose $d_t$ versus $d'_t$ in period $t$ given covariates $X$ and the previous treatment history $\mathbf{D}_{t-1}$ is identified analogously as expression \eqref{DiDobs}:
\begin{eqnarray}\label{DiDobs2}
&&E[Y_t(d_t)-Y_t(d'_t)|D_t=d_t,T=t,\mathbf{D}_{t-1}, X]\\
&=&E[Y_t(d_t)|D_t=d_t,T=t,\mathbf{D}_{t-1},X]-E[Y_{t-1}(d_{t-1})|D_t=d_t,T=t-1,\mathbf{D}_{t-1},X]\notag\\
&-&\{E[Y_t(d_t')|D_t=d'_t,T=t,\mathbf{D}_{t-1},X]-E[Y_{t-1}(d_{t-1})|D_t=d_t,T=t-1,\mathbf{D}_{t-1},X]\}.\notag
\end{eqnarray}
Integrating the conditional ATET over the distribution of $X$ and $\mathbf{D}_{t-1}$ yields the ATET at a treatment dose $D_t = d_t$ in that period:
\begin{eqnarray}\label{DiDidentpostreg2}
\Delta_{d_t,d_t', t}&=&E[  \mu_{d_t}(t,\mathbf{D}_{t-1},X)-\mu_{d_t}(t-1,\mathbf{D}_{t-1},X)\\ 
&-& (\mu_{d'_t}(t,\mathbf{D}_{t-1},X)-\mu_{d'_t}(t-1,\mathbf{D}_{t-1},X))|D_t=d_t, T=t ],\notag
\end{eqnarray}
where $\mu_{d_t}(t',\mathbf{D}_{t-1},x)=E[Y_T|D_t=d_t,T=t',\mathbf{D}_{t-1}=\mathbf{D}_{t-1},X=x]$ is the conditional mean outcome in period $t'$ given the treatment received in period $t$, the previous treatment history up to period $t-1$, and the covariates. We note that for the case that $d_t'=d_{t-1}''$, the ATET corresponds to a causal effect of switching from the previous treatment dose $d_t'$ to a higher treatment dose $d_t$ among switchers. As a comparison, \cite{dechaisemartin2023differenceindifferences} consider average effects among switchers across different treatment margins, rather than exclusively switching to dose $d_t$.

As an alternative to equation \eqref{DiDidentpostreg2}, the ATET may be formalized by the following, equivalent DR expression:
\begin{align}\label{DiDidentDRcont2}
&\Delta_{d_t,d_t', t}= \notag\\
&\lim_{h \rightarrow 0} E\left[ \frac{  \omega(D_t;d_t, h) \cdot I\{T=t\}\cdot [Y_T-\mu_{d_{t}}(t-1,\mathbf{D}_{t-1},X)-\mu_{d'_t}(t,\mathbf{D}_{t-1},X)+\mu_{d'_t}(t-1,\mathbf{D}_{t-1},X) ] }{ \Pi_{d_t,t}}\right.\notag\\ 
&- \frac{\omega(D_t;d_t, h) \cdot I\{T=t-1\}\cdot \rho_{d_t,t}(\mathbf{D}_{t-1},X)\cdot \epsilon_{d_t,t-1,\mathbf{D}_{t-1},X}}{ \rho_{d_t,t-1}(\mathbf{D}_{t-1},X)\cdot \Pi_{d_t,t}}\\
&- \left(\frac{\omega(D_t;d'_t, h)\cdot  I\{T=t\}\cdot \rho_{d_t,t}(\mathbf{D}_{t-1},X)\cdot \epsilon_{d'_t,t,\mathbf{D}_{t-1},X}}{ \rho_{d'_t,t}(\mathbf{D}_{t-1},X) \cdot \Pi_{d_t,t}}\right. \notag\\
&- \left.\left.\frac{\omega(D_t;d'_t, h)\cdot I\{T=t-1\}\cdot \rho_{d_t,t}(\mathbf{D}_{t-1},X)\cdot \epsilon_{d'_t,t-1,\mathbf{D}_{t-1},X}}{ \rho_{d_t',t-1}(\mathbf{D}_{t-1},X)\cdot \Pi_{d_t,t}}\right)  \right],\notag
\end{align}
where $\epsilon_{{d_t},t,\mathbf{D}_{t-1},x}=(Y_T-\mu_{d_t}(t,\mathbf{D}_{t-1},x))$ denotes a regression residual. 

As a further modification in our DiD framework, we might also be interested in the ATET of a lagged treatment dose on outcomes in later periods, rather than only evaluating the the instantaneous impact of a treatment dose on an outcome in the same period. Hence, let us denote by $s\geq 1$ the time lag between the treatment and the outcome period, which permits defining the ATET of a treatment dose in period $t-s$ on an outcome in period $t$: 
\begin{align}\label{ATElagged}\Delta_{d_{t-s},d'_{t-s}, t}=E[Y_t(d_{t-s})-Y_t(d'_{t-s})|D_{t-s}=d_{t-s},T=t],\end{align} with time lag $s\geq 1$. 
\noindent We note that the ATET $\Delta_{d_{t-s},d'_{t-s}, t}$ comprises both the direct effect effect of the treatment dose in period $t-s$ on the outcome in $t$, as well as the indirect (or mediated) effect of the treatment dose in period $t-s$ on the treatments in later periods, that in turn also affect the outcome in period $t-s$. $\Delta_{d_{t-s},d'_{t-s}, t}$ is identified when replacing Assumption \ref{ass5} by the following parallel trend assumption:
\begin{assumption} {\bf (Conditional parallel trends with time lags):}\label{ass8}\\
$E[Y_t(d_{t-s}')-Y_{t-s-1}(d_{t-s-1}'')|D_{t-s}=d_{t-s},T=t, \mathbf{D}_{t-s-1}=\mathbf{d}_{t-s-1}'',X] \\ = E[Y_t(d_{t-s}')-Y_{t-s-1}(d_{t-s-1}'')|D_{t-s}=d_{t-s}', T=t, \mathbf{D}_{t-s-1}=\mathbf{d}_{t-s-1}'',X]$ for  $d_{t-s}>d_{t-s}'\geq 0$ in the support of $D_{t-s}$, $\mathbf{d}_{t-s-1}''\geq \mathbf{0}$ in the support of $\mathbf{D}_{t-s-1}$, $t>1$ in the support of $T$, and $1 \leq s < t$.
\end{assumption}

Assumption \ref{ass8} is stronger than Assumption \ref{ass5}, because it requires that changes in the treatment dose after period $t-s$ up to period $t$ do not evolve endogenously. This means that unobserved confounders jointly affecting the treatment doses in period $t-s$ and subsequent $s$ periods are ruled out. Otherwise, the parallel trend assumption would generally fail, as the outcome (trend) is typically also a function of the treatments after period $t-s$. Therefore, the presence of unobserved confounders jointly affecting the treatment doses in period $t-s$ and subsequent periods would ultimately imply confounding of treatment $D_{t-s}$ and outcome $Y_t$. In contrast, Assumption \ref{ass5} does not put restrictions on confounders of the treatment dose over time, with the caveat that the ATET may only be assessed on outcomes in the same period as the treatment dose to be evaluated, but not on outcomes in later periods.

To better see the implications of Assumption \ref{ass8} in terms of ruling out confounders that jointly affect treatments in different periods, consider the following example. 
Suppose the outcome in time period $T$ is a (possibly unknown) function, denoted by $\mathcal{F}_Y$, of the treatments in periods 1 and 2, $D_1$ and $D_2$, covariates $X$, and time $T$ (pre-treatment histories are omitted for simplicity), together with an additively separable, time-constant function $\mathcal{F}_U$ of time-invariant unobservables:
\[
Y_T = \mathcal{F}_Y(D_1, D_2, X, T) + \mathcal{F}_U(U).
\]
Further, assume that the first treatment is a function of $X$ and $U$,
\[
D_1 = \mathcal{F}_{D_1}(X, U),
\]
and that the second treatment depends on the first treatment and $X$,
\[
D_2 = \mathcal{F}_{D_2}(D_1, X).
\]
We may add random, time-varying, and additively separable error terms to the models for $Y_T$, $D_1$, and $D_2$, but these are omitted here for simplicity.

Assume we are interested in the ATET of the treatment in period $T=1$ on the outcome in period $T=2$. 
Differencing average potential outcomes under treatment value $d_1'$ between period $T=2$ and the pre-treatment period $T=0$, conditional on $X=x$, eliminates the time-constant component $\mathcal{F}_U(U)$ and yields
\begin{align}\label{D1D2noprob}
E[Y_2(d'_1) - Y_0(d'_1) \mid X=x] 
= E[\mathcal{F}_Y(d'_1, D_2(d'), x, 2) - \mathcal{F}_Y(d', D_2(d'), x, 0)],
\end{align}
where $D_2(d_1')$ denotes the potential second-period treatment under a first-period dose $d_1'$.

Under the maintained model for $D_2$ (i.e., $D_2 = \mathcal{F}_{D_2}(D_1, X)$), the potential second-period treatment is $D_2(d_1') = \mathcal{F}_{D_2}(d_1', X)$ and hence does not depend on $U$. 
Consequently, $U$, while affecting $D_1$, does not affect the quantity on the right-hand side of \eqref{D1D2noprob}; therefore, Assumption \ref{ass8} is satisfied in this setup.

By contrast, if the second treatment also depends on $U$, say
\[
D_2 = \mathcal{F}_{D_2}(D_1, X, U),
\]
then the potential second-period treatment becomes $D_2(d_1') = \mathcal{F}_{D_2}(d_1', X, U)$, and $U$ enters the right-hand side of the differenced expression. In that case
\begin{align}\label{D1D2}
E[Y_2(d'_1) - Y_0(d'_1) \mid X=x]  = E[\mathcal{F}_Y(d_1', \mathcal{F}_{D_2}(d_1', x, U), x, 2) - \mathcal{F}_Y(d', \mathcal{F}_{D_2}(d_1', x, U), x, 0)],
\end{align}
so $U$ affects the mean potential outcome difference via the second treatment and also through the first treatment (since $D_1$ depends on $U$). Hence $U$ acts as a confounder for the effect of the first-period treatment on the differenced outcome, and Assumption \ref{ass8} is violated. In summary, Assumption \ref{ass8} rules out precisely such scenarios in which unobservables affect both the early-period treatment $D_{t-s}$ and subsequent treatments that also affect the outcome.

To identify the ATET under Assumptions \ref{ass4}, \ref{ass6}, and \ref{ass7} (when considering $t-s$ rather than $t$ as treatment period), as well as Assumption \ref{ass8} using a DR expression, we need to appropriately modify equation \eqref{DiDidentDRcont2} to include time lags when defining the pre- and post-treatment periods of interest: 
\begin{align}\label{DiDidentDRcont3}
&\Delta_{d_{t-s},d'_{t-s}, t} = \lim_{h \rightarrow 0} E\left[ \left\{ \frac{  \omega(D_{t-s};d_{t-s}, h) \cdot I\{T=t\}}{ \Pi_{d_{t-s},t}}\right.\right.\notag\\
&\times [Y_T-\mu_{d_{t-s}}(t-s-1,\mathbf{D}_{t-s-1},X)-\mu_{d'_{t-s}}(t,\mathbf{D}_{t-s-1},X)+\mu_{d'_{t-s}}(t-s-1,\mathbf{D}_{t-s-1},X) ]\notag\\
&- \frac{\omega(D_{t-s};d_{t-s}, h) \cdot I\{T=t-s-1\}\cdot \rho_{d_{t-s},t}(\mathbf{D}_{t-s-1},X)\cdot\epsilon_{d_{t-s},t-s-1,\mathbf{D}_{t-s-1},X}}{ \rho_{d_{t-s},t-s-1}(\mathbf{D}_{t-s-1},X)\cdot \Pi_{d_{t-s},t}}\notag\\
&- \left(\frac{\omega(D_{t-s};d'_{t-s}, h)\cdot  I\{T=t\}\cdot \rho_{d_{t-s},t}(\mathbf{D}_{t-s-1},X)\cdot\epsilon_{d'_{t-s},t,\mathbf{D}_{t-s-1},X}}{ \rho_{d'_{t-s},t}(\mathbf{D}_{t-s-1},X) \cdot \Pi_{d_{t-s},t}} \right.\\
&-\left.\left.\left.\frac{\omega(D_{t-s};d'_{t-s}, h)\cdot I\{T=t-s-1\}\cdot \rho_{d_{t-s},t}(\mathbf{D}_{t-s-1},X)\cdot\epsilon_{d'_{t-s},t-s-1,\mathbf{D}_{t-s-1},X}}{ \rho_{d_{t-s}',t-s-1}(\mathbf{D}_{t-s-1},X)\cdot \Pi_{d_{t-s},t}}\right) \right\} \right].\notag
\end{align}
Appendix \ref{Neymanrepeated} demonstrates that DR expressions \eqref{DiDidentDRcont}, \eqref{DiDidentDRcont2}, and \eqref{DiDidentDRcont3} satisfy Neyman orthogonality as the kernel bandwidth goes to zero. Therefore, these expressions can be employed to construct DML estimators under certain regularity conditions outlined in \cite{Chetal2018}. 

We note that the identification of $\Delta_{d_{t-s},d'_{t-s}, t}$ under parallel trends as imposed in Assumption \ref{ass8} 
is distinct from the scenario considered in \cite{de2024difference}, where the counterfactual is defined in terms of a fixed 
treatment dose $d'$ over all post-treatment periods $t-s$ to $t$, that is, conditional on being an non-switcher in terms of treatment doses. This scenario allows considering the following ATET:
\begin{align}\label{ATElagged2}
\Delta_{d_{t-s},\mathbf{d'}_{t-s:t}, t}
&=E[Y_t(d_{t-s})-Y_t(d'_{t-s},...,d'_{t})\mid D_{t-s}=d_{t-s},T=t]\\
&=E[Y_t(d_{t-s})-Y_t(\mathbf{d'}_{t-s:t})\mid D_{t-s}=d_{t-s},T=t],\notag
\end{align} 
where $Y_t(\mathbf{d'}_{t-s:t})$ denotes the lagged post-treatment counterfactual outcome 
under the sequence of fixed treatment doses $\mathbf{d'}_{t-s:t} = (d'_{t-s}, \ldots, d'_t)$. Identification of the ATET in \eqref{ATElagged2} requires yet another parallel trend  assumption, different from Assumption \ref{ass8} and more closely related to the one considered in \cite{de2024difference}:

\begin{assumption}{\bf (Conditional parallel trends under a fixed treatment sequence):}\label{ass8a}\\
$E\left[\,Y_t(d'_{t-s}, \ldots, d'_t) - Y_{t-s-1}(d''_{t-s-1}) \mid D_{t-s}=d_{t-s}, T=t, 
\mathbf{D}_{t-s-1}=\mathbf{d}''_{t-s-1}, X \right] \\
=
E\left[\,Y_t(d'_{t-s}, \ldots, d'_t) - Y_{t-s-1}(d''_{t-s-1}) \mid D_{t-s}=d'_{t-s}, T=t, 
\mathbf{D}_{t-s-1}=\mathbf{d}''_{t-s-1}, X \right]$
for $d_{t-s}>(d_{t-s}',...,d'_t)\geq 0$ in the support of $(D_{t-s},...,D_t)$, 
$\mathbf{d}_{t-s-1}''\geq \mathbf{0}$ in the support of $\mathbf{D}_{t-s-1}$, 
$t>1$ in the support of $T$, and $1 \leq s < t$.
\end{assumption}

Assumption \ref{ass8a} requires that, conditional on covariates $X$ and the pre-period treatment history $\mathbf{D}_{t-s-1}$, 
the mean trend in potential outcomes under the sequence of fixed treatment doses $\mathbf{d}'_{t-s:t}$ 
is the same for units whose realized treatment at time $t-s$ equals $d_{t-s}$ and for those with $d'_{t-s}$. 
In contrast to Assumption \ref{ass8}, Assumption \ref{ass8a} allows for unobserved factors that jointly affect treatments in different time periods (provided that these factors do not also affect the differences in mean potential outcomes between the pre- and post-treatment periods). In the example given in equation \eqref{D1D2}, which entails the violation of Assumption \ref{ass8}, Assumption \ref{ass8a} is satisfied because, 
conditional on both the first- and second-period treatments, time-invariant unobserved confounders are differenced out from the mean potential outcome equations:
\begin{align}\label{D1D2}
E[Y_2(d'_1, d'_2) - Y_0(d'_1,d'_2) \mid X=x]  
= E[\mathcal{F}_1(d'_1, d'_2, x, 2) - \mathcal{F}_1(d'_1, d'_2, x, 0)].
\end{align}

In addition to a modification of the parallel trend assumption, the nonparametric identification of the ATET in \eqref{ATElagged2}  requires strengthening the previous common support condition to hold with respect to the conditional density for the sequence:
\begin{assumption}{\bf (Common support under a fixed treatment sequence):}\label{ass7a}\\
$f(D_{t-s} = d_{t-s},\, T = t \mid \mathbf{D}_{t-s-1}, X,\, (D_{t-s}, T) 
\in \{(d^*_t, t'), (d_{t-s}, t)\}) < 1
\\ \text{for } d_{t-s} > d'_{t-s} 
\text{ in the support of } D_{t-s},\;
t > 0 \text{ in the support of } T,\;
\text{and } (d^*_t, t') \in \{(d_{t-s}, t-s-1), (\mathbf{d}'_{t-s:t}, t), 
(\mathbf{d}'_{t-s:t}, t-s-1)\}.$
\end{assumption}
This common support condition becomes more stringent as the number of post-treatment periods increases.

Under Assumptions \ref{ass4}, \ref{ass6}, \ref{ass8a}, and \ref{ass7a}, the ATET $\Delta_{d_{t-s},\mathbf{d'}_{t-s:t}, t}$ is identified by appropriately modifying DR expression \eqref{DiDidentDRcont3}. To account for a fixed treatment sequence $\mathbf{d}'_{t-s:t}$, we define the product kernel
\begin{align}\label{prodkernel}
\omega(\mathbf{D}_{t-s:t}; \mathbf{d}'_{t-s:t}, \mathbf{h})
= \prod_{r=0}^{s} \frac{1}{h_r} \, \mathcal{K}\!\left(\frac{D_{t-s+r} - d'_{t-s+r}}{h_r}\right),
\end{align}
where $\mathbf{h} = (h_0, \ldots, h_s)$ is a vector of bandwidths and $\mathcal{K}$ is a univariate kernel. Furthermore, we denote by 
$\mu_{\mathbf{d}'_{t-s:t}}(t, \mathbf{D}_{t-s-1}, X)
= E[Y_t \mid \mathbf{D}'_{t-s:t}=\mathbf{d}'_{t-s:t}, \mathbf{D}_{t-s-1}, X]$ the conditional mean outcome under the treatment sequence, and by 
$\epsilon_{\mathbf{d}'_{t-s:t},t,\mathbf{D}_{t-s-1},X} 
= Y_t - \mu_{\mathbf{d}'_{t-s:t}}(t, \mathbf{D}_{t-s-1}, X)$ the corresponding residual. 
Finally, the conditional density of the treatment sequence is given by $
\rho_{\mathbf{d}'_{t-s:t},t}(\mathbf{D}_{t-s-1},X)= f(\mathbf{D}'_{t-s:t}=\mathbf{d}'_{t-s:t} 
\mid \mathbf{D}_{t-s-1}, X, T=t)$. Then, the following DR expression 
identifies the ATET defined in \eqref{ATElagged2}:
\begin{align}\label{DiDidentDRcont4}
&\Delta_{d_{t-s},\mathbf{d'}_{t-s:t}, t} = \lim_{h \rightarrow 0} E\left[ \left\{ 
\frac{  \omega(D_{t-s};d_{t-s}, h) \cdot I\{T=t\}}{ \Pi_{d_{t-s},t}}\right.\right.\notag\\
&\times [Y_T-\mu_{d_{t-s}}(t-s-1,\mathbf{D}_{t-s-1},X)-\mu_{\mathbf{d}'_{t-s:t}}(t,\mathbf{D}_{t-s-1},X)
+\mu_{\mathbf{d}'_{t-s:t}}(t-s-1,\mathbf{D}_{t-s-1},X) ]\notag\\
&- \frac{\omega(D_{t-s};d_{t-s}, h) \cdot I\{T=t-s-1\}\cdot 
\rho_{d_{t-s},t}(\mathbf{D}_{t-s-1},X)\cdot
\epsilon_{d_{t-s},t-s-1,\mathbf{D}_{t-s-1},X}}{ 
\rho_{d_{t-s},t-s-1}(\mathbf{D}_{t-s-1},X)\cdot \Pi_{d_{t-s},t}}\notag\\
&- \left(\frac{\omega(\mathbf{D}_{t-s:t}; \mathbf{d}'_{t-s:t}, \mathbf{h})\cdot  
I\{T=t\}\cdot \rho_{d_{t-s},t}(\mathbf{D}_{t-s-1},X)\cdot
\epsilon_{\mathbf{d}'_{t-s:t},t,\mathbf{D}_{t-s-1},X}}{ 
\rho_{\mathbf{d}'_{t-s:t},t}(\mathbf{D}_{t-s-1},X) \cdot \Pi_{d_{t-s},t}} \right.\\
&-\left.\left.\left.\frac{\omega(\mathbf{D}_{t-s:t}; \mathbf{d}'_{t-s:t}, \mathbf{h})\cdot 
I\{T=t-s-1\}\cdot \rho_{d_{t-s},t}(\mathbf{D}_{t-s-1},X)\cdot
\epsilon_{\mathbf{d}'_{t-s:t},t-s-1,\mathbf{D}_{t-s-1},X}}{ 
\rho_{\mathbf{d}'_{t-s:t},t-s-1}(\mathbf{D}_{t-s-1},X)\cdot 
\Pi_{d_{t-s},t}}\right) \right\} \right].\notag
\end{align}
We note that in empircal applications, the use of product kernels in $(s+1)$ dimensions (related to the number of post-treatment periods) introduces the usual curse of dimensionality, since the effective sample size scales with $\prod_{r=0}^s h_r$.

\section{Identification in panel data}\label{paneldata}

This section discusses identification in panel data under the identifying assumptions proposed in the previous section. We first consider the identification of $\Delta_{d_t, d_t', t}$, i.e., the ATET of treatment dose $d_t$ versus $d_t'$ in period $t$ (net of direct effects of previous treatment doses) as defined in equation \eqref{ATEsameperiod}. As panel data permits taking outcome differences within subjects over time, the identification result in equation \eqref{DiDidentpostreg2} simplifies as follows:

\begin{eqnarray}\label{DiDidentpostreg2modified}
\Delta_{d_t,d_t', t}
&=&E[E[Y_{t}-Y_{t-1}|D_t=d_t,\mathbf{D}_{t-1}=\mathbf{D}_{t-1},X=x]\notag\\
&-& \{E[Y_{t}-Y_{t-1}|D_t=d_t',\mathbf{D}_{t-1}=\mathbf{D}_{t-1},X=x]\}|D_t=d_t].
\end{eqnarray}

It is worth noticing that dropping $T=t$ from the conditioning follows from the fact that, in panel data, the same subjects are observed in periods $t$ and $t-1$, such that distributions of $\mathbf{D}_{t-1}$ and $X$ are constant across periods by design. To ease notation, we denote the conditional mean differences in outcomes by $m_{d_t}(t,\mathbf{D}_{t-1},X)=E[Y_{t}-Y_{t-1}|D_t=d_t,\mathbf{D}_{t-1},X]$. Moreover, we use $P_{d_t}=f(D_t=d_t)$ and $p_{d_t}(\mathbf{D}_{t-1},X)=f(D_t=d_t|\mathbf{D}_{t-1},X)$ to denote the (conditional) density functions of treatment $D_t=d_t$. Specifically, it can be shown that $P_{d_t}=\lim_{h \rightarrow 0} E[\omega(D;d_t,h)]$ and $p_{d_t}(\mathbf{D}_{t-1},X)$$=\lim_{h \rightarrow 0} E[\omega(D;d_t,h)|\mathbf{D}_{t-1},X]$ hold. We propose an DR expression that is a straightforward modification of the one provided in equation (3.8) of \cite{zhang2024}. The key differences are that the treatment dose in the control group is non-zero, thus involving kernel functions with respect to both $d_t'$ and $d_t$, and that the conditioning set additionally includes the treatment history $\mathbf{D}_{t-1}$ due to the multiple-period setup. Our proposed DR expression for equation \eqref{DiDidentpostreg2modified} for ATET identification is given by:
\begin{align}\label{DiDidentDRcont2modified}
\Delta_{d_t,d_t', t} &= \lim_{h \rightarrow 0} E\left[ \frac{  \omega(D_t;d_t, h) \cdot [Y_t-Y_{t-1}-m_{d_{t}'}(t,\mathbf{D}_{t-1},X)] }{ P_{d_t}}\right.\notag\\ 
&- \left.\frac{\omega(D_t;d_t', h)\cdot  p_{d_t}(\mathbf{D}_{t-1},X)\cdot [Y_t-Y_{t-1}-m_{d_{t}'}(t,\mathbf{D}_{t-1},X)] }{p_{d_t'}(\mathbf{D}_{t-1},X) \cdot P_{d_t}}\right].
\end{align} 

To see the equivalence of equations \eqref{DiDidentDRcont2modified} and \eqref{DiDidentpostreg2modified}, we note that by the law of iterated expectation,
\begin{align*}
&\lim_{h \rightarrow 0} E\left[ \frac{  \omega(D_t;d_t, h) \cdot [Y_t-Y_{t-1}-m_{d_{t}'}(t,\mathbf{D}_{t-1},X)] }{ P_{d_t}}\right]\\
&= E[Y_t-Y_{t-1}-m_{d_{t}'}(t,\mathbf{D}_{t-1},X)|D_t=d_t]= E[m_{d_{t}}(t,\mathbf{D}_{t-1},X)-m_{d_{t}'}(t,\mathbf{D}_{t-1},X)|D_t=d_t],
\end{align*} 
which corresponds to the ATET of interest. In addition, again by the law of iterated expectations, we have that
\begin{align*}
&E\left[ \frac{\omega(D_t;d_t', h)\cdot  p_{d_t}(\mathbf{D}_{t-1},X) } {p_{d_t'}(\mathbf{D}_{t-1},X) \cdot P_{d_t}} \cdot [Y_t-Y_{t-1}-m_{d_{t}'}(t,\mathbf{D}_{t-1},X)] \right]\\
&= E\left[ \frac{  p_{d_t}(\mathbf{D}_{t-1},X) }{ P_{d_t}} \cdot E\left[\frac{\omega(D_t;d_t', h)\cdot [Y_t-Y_{t-1}-m_{d_{t}'}(t,\mathbf{D}_{t-1},X)]}{p_{d_t'}(\mathbf{D}_{t-1},X)}\Big| X \right] \right]\\
&= E\left[  m_{d_{t}'}(t,\mathbf{D}_{t-1},X)-m_{d_{t}'}(t,\mathbf{D}_{t-1},X) | D_t=d_t\right]=0,
\end{align*} 
which demonstrates that both equations \eqref{DiDidentDRcont2modified} and \eqref{DiDidentpostreg2modified} identify the ATET under Assumptions \ref{ass4} through \ref{ass7} (without conditioning on $T=t$).

Next, we consider evaluating $\Delta_{d_{t-s},d'_{t-s}, t}$, the ATET of receiving the treatment dose $d$ versus $d'$ in period $t-s$ on a later outcome in period $t$, as defined in equation \eqref{ATElagged}. Under Assumptions \ref{ass4}, \ref{ass6}, \ref{ass7}, and \ref{ass8}, a DR expression in the panel setting with time lags, analogous to \eqref{DiDidentDRcont2} for repeated cross sections, is given by the following expression:
\begin{align}\label{DiDidentDRcont3modified}
\Delta_{d_{t-s},d'_{t-s}, t} &= \lim_{h \rightarrow 0} E\left[ \frac{  \omega(D_{t-s};d_{t-s}, h) \cdot [Y_t-Y_{t-s-1}-m_{d_{t-s}'}(t,\mathbf{D}_{t-s-1},X)] }{ P_{d_{t-s}}}\right.\\ 
&- \left.\frac{\omega(D_{t-s};d_{t-s}', h)\cdot  p_{d_{t-s}}(\mathbf{D}_{t-s-1},X)\cdot [Y_t-Y_{t-s-1}-m_{d_{t-s}'}(t,\mathbf{D}_{t-s-1},X)] }{p_{d_{t-s}'}(\mathbf{D}_{t-s-1},X) \cdot P_{d_{t-s}}}\right],\notag
\end{align} 
where $m_{d_{t-s}'}(t,\mathbf{D}_{t-s-1},X)=E[Y_{t}-Y_{t-s-1}|D_{t-s}=d_{t-s},\mathbf{D}_{t-s-1},X]$. Appendix \ref{Neymanrepeated} demonstrates that DR expressions \eqref{DiDidentDRcont2modified} and \eqref{DiDidentDRcont3modified} satisfy Neyman orthogonality. 

Finally, when assessing the ATET $\Delta_{d_{t-s},\mathbf{d'}_{t-s:t}, t}$ under the fixed treatment sequence defined in equation \eqref{ATElagged2}, equation \eqref{DiDidentDRcont3modified} needs to be modified by making use of the product kernel defined in equation \eqref{prodkernel}, the conditional mean outcome differences under fixed treatment sequences,
$m_{\mathbf{d}_{t-s:t}'}(t,\mathbf{D}_{t-1},X)=E[Y_{t}-Y_{t-1} \mid \mathbf{D}_{t-s:t}'=\mathbf{d}_{t-s:t}',\mathbf{D}_{t-1},X]$, 
and the corresponding conditional densities of fixed treatment sequences,
$p_{\mathbf{d}_{t-s:t}'}(\mathbf{D}_{t-1},X)=f(\mathbf{D}_{t-s:t}'=\mathbf{d}_{t-s:t}' \mid \mathbf{D}_{t-1},X)$. 
Under Assumptions \ref{ass4}, \ref{ass6}, \ref{ass8a}, and \ref{ass7a}, the following expression identifies the ATET in panel data:
\begin{align}\label{DiDidentDRcont4modified}
\Delta_{d_{t-s},\mathbf{d}_{t-s:t}', t} &= \lim_{h \rightarrow 0} E\left[ \frac{  \omega(D_{t-s};d_{t-s}, h) \cdot [Y_t-Y_{t-s-1}-m_{\mathbf{d}_{t-s:t}'}(t,\mathbf{D}_{t-s-1},X)] }{ P_{d_{t-s}}}\right.\\ 
&- \left.\frac{\omega(\mathbf{D}_{t-s:t}; \mathbf{d}'_{t-s:t}, \mathbf{h})\cdot  p_{d_{t-s}}(\mathbf{D}_{t-s-1},X)\cdot [Y_t-Y_{t-s-1}-m_{\mathbf{d}_{t-s:t}'}(t,\mathbf{D}_{t-s-1},X)] }{p_{\mathbf{d}_{t-s:t}'}(\mathbf{D}_{t-s-1},X) \cdot P_{d_{t-s}}}\right].\notag
\end{align}

\section{Estimation}
\label{sec:meth}

In this section, we propose DiD estimators based on the DML framework as discussed in \cite{Chetal2018}, using the sample analogs of the DR identification results in sections \ref{sec:id} and \ref{paneldata}. We first consider estimating the ATET $\Delta_{d_t,d_t', t}$ as defined in equation \eqref{ATEsameperiod} with the sample version of equation \eqref{DiDidentDRcont2}. Thus, let $\mathcal{W} = \{W_i|1\leq i \leq n\}$ with $W_i = (Y_{i,T}, D_{i,t}, \mathbf{D}_{i,t-1} X_i, T_i)$ for $i=1,\ldots, n$ denote the set of observations in an i.i.d.\ sample of size $n$.
Moreover, let 
\begin{align} 
  \eta =&  \{ \mu_{d_{t}}(t-1,\mathbf{D}_{t-1},X),\mu_{d'_t}(t,\mathbf{D}_{t-1},X),\mu_{d'_t}(t-1,\mathbf{D}_{t-1},X),\notag\\ 
  &\rho_{d_t,t}(\mathbf{D}_{t-1},X), \rho_{d_t,t-1}(\mathbf{D}_{t-1},X),\rho_{d'_t,t}(\mathbf{D}_{t-1},X),\hat\rho_{d'_t,t-1}(\mathbf{D}_{t-1},X)\}\notag 
  \end{align}
  denote the set of nuisance parameters, i.e., the conditional mean outcomes and treatment densities. Their respective estimates are denoted by 
  \begin{align} 
  \hat{\eta} =&  \{\hat\mu_{d_{t}}(t-1,\mathbf{D}_{t-1},X),\hat\mu_{d'_t}(t,\mathbf{D}_{t-1},X),\hat\mu_{d'_t}(t-1,\mathbf{D}_{t-1},X),\notag\\ 
  & \hat\rho_{d_t,t}(\mathbf{D}_{t-1},X), \hat\rho_{d_t,t-1}(\mathbf{D}_{t-1},X),\hat\rho_{d'_t,t}(\mathbf{D}_{t-1},X),\hat\rho_{d'_t,t-1}(\mathbf{D}_{t-1},X)\}.\notag
  \end{align}
The following algorithm formally sumarizes the estimation procedure for $\Delta_{d_t,d_t',t}$:   
\vspace{15pt}\newline
  \textbf{DML algorithm:} 
  \begin{enumerate}
  	\item Split $\mathcal{W}$ in $ K $ subsamples. For each subsample $k$, let $n_k$ denote its size, $\mathcal{W}_k$ the set of observations in the sample and $\mathcal{W}_k^{C}$ the complement set of all observations not in $\mathcal{W}_k$.
  	\item For each $k$, use $\mathcal{W}_k^{C}$ to estimate the model parameters of the nuisance parameters $\eta$ by machine learning in order to predict these nuisance parameters in $\mathcal{W}_k$, where the predictions are denoted by $\hat{\eta}^k$.
  	\item Stack the fold-specific estimates of the nuisance parameters $\hat{\eta}^1,...,\hat{\eta}^K$ to generate a matrix of nuisance estimates $\hat{\eta}$ for the entire sample: $\hat{\eta} = \begin{pmatrix} \hat{\eta}^1 \\ \vdots \\ \hat{\eta}^K \end{pmatrix}$
  	\item Plug the nuisance parameters into the normalized sample analog of equation \eqref{DiDidentDRcont2} to obtain an estimate of the ATET using bandwidth $h$, denoted by $\hat\Delta_{d_t,d_t', t}^h$:

\begin{flalign}
\hat\Delta_{d_t,d_t',t}^h& = \sum_{i=1}^{n} \left[\biggl(\sum_{j=1}^n \omega(D_{j,t};d_t,h) \cdot I\{T_j=t\}\biggr)^{-1} \cdot (\omega(D_{i,t};d_t, h) \cdot I\{T_i=t\} \right. \notag \\
&\cdot \left[Y_{i,T}-\hat\mu_{d_{t}}(t-1,\mathbf{D}_{i,t-1},X_i)-\hat\mu_{d'_t}(t,\mathbf{D}_{i,t-1},X_i)+\hat\mu_{d'_t}(t-1,\mathbf{D}_{i,t-1},X_i)\right])\notag \\
&- \frac{ \omega(D_{i,t};d_t, h) \cdot I\{T_i=t-1\}\cdot \hat\rho_{d_t,t}(\mathbf{D}_{i,t-1},X_i)\cdot \hat\epsilon_{i,d_t,t-1,\mathbf{D}_{i,t-1},X_i}/ \hat\rho_{d_t,t-1}(\mathbf{D}_{i,t-1},X)}{
(\sum_{j=1}^n \omega(D_{j,t};d_t, h) \cdot I\{T_j=t-1\})\cdot \hat\rho_{d_t,t}(\mathbf{D}_{i,t-1},X_i)/ \hat\rho_{d_t,t-1}(\mathbf{D}_{i,t-1},X)}\notag \\
&- \left(\frac{\omega(D_{i,t};d'_t, h)\cdot  I\{T_i=t\}\cdot \hat\rho_{d_t,t}(\mathbf{D}_{i,t-1},X)\cdot \hat\epsilon_{i,d'_t,t,\mathbf{D}_{i,t-1},X_i}/\hat\rho_{d'_t,t}(\mathbf{D}_{i,t-1},X_i)}{(\sum_{j=1}^n \omega(D_{j,t};d'_t, h)\cdot  I\{T_j=t\})\cdot \hat\rho_{d_t,t}(\mathbf{D}_{i,t-1},X_i)/\hat\rho_{d'_t,t}(\mathbf{D}_{i,t-1},X_i)} \right. \notag \\
&- \left.\left.\frac{\omega(D_{i,t};d'_t, h)\cdot I\{T_i=t-1\}\cdot \hat\rho_{d_t,t}(\mathbf{D}_{i,t-1},X)\cdot \hat\epsilon_{i,d'_t,t-1,\mathbf{D}_{i,t-1},X_i}/\hat\rho_{d_t',t-1}(\mathbf{D}_{i,t-1},X_i)}{(\sum_{j=1}^n \omega(D_{j,t};d'_t, h)\cdot I\{T_j=t-1\})\cdot \hat\rho_{d_t,t}(\mathbf{D}_{i,t-1},X_i)/\hat\rho_{d_t',t-1}(\mathbf{D}_{i,t-1},X_i)}\right) \right],\notag
\end{flalign}
where $\hat\epsilon_{i,{d_t},t,\mathbf{D}_{i,t-1},X_i}=(Y_{i,T}-\hat\mu_{d_t}(t,\mathbf{D}_{i,t-1},X_i))$ denotes the estimated regression residual. 
   \vspace{5pt}\newline
  \end{enumerate}

For estimation in panel data, we assume an i.i.d. sample of $n$ subjects with $W_i = (Y_{i,t}, Y_{i,t-1}, D_{i,t}, \mathbf{D}_{i,t-1} X_i)$ for $i=1,\ldots, n$. Then, the nuisance parameter vector corresponds to 
\begin{align*} 
 \eta =&  \{ m_{d'_t}(t,\mathbf{D}_{t-1},X),  p_{d_t}(\mathbf{D}_{t-1},X),  p_{d'_t}(\mathbf{D}_{t-1},X)\},
    \end{align*}
    and its estimate to 
\begin{align*} 
  \hat{\eta} =&  \{\hat m_{d'_t}(t,\mathbf{D}_{t-1},X), \hat p_{d_t}(\mathbf{D}_{t-1},X), \hat p_{d'_t}(\mathbf{D}_{t-1},X)\}.
  \end{align*}
  ATET estimation proceeds analogously as outlined in the algorithm above, although using the sample analog of equation \eqref{DiDidentDRcont2modified} in step 4 of ATET estimation:
   \begin{align} 
\hat\Delta_{d_t,d_t', t}^h&=\sum_{i=1}^{n}\left[ \frac{ \omega(D_{i,t};d_t, h) \cdot [Y_{i,t}-Y_{i,t-1}-\hat m_{d'_t}(t,t-1,\mathbf{D}_{i,t-1},X_i) ] }{  \sum_{j=1}^n \omega(D_{j,t};d_t, h) }\right.\notag\\ 
&\left.- \frac{\omega(D_{i,t};d'_t, h)\cdot \hat\rho_{d_t}(\mathbf{D}_{i,t-1},X)\cdot[Y_{i,t}-Y_{i,t-1}-\hat m_{d'_t}(t,t-1,\mathbf{D}_{i,t-1},X_i) ]/\hat\rho_{d'_t}(\mathbf{D}_{i,t-1},X_i)}{(\sum_{j=1}^n \omega(D_{j,t};d'_t, h))\cdot \hat\rho_{d_t}(\mathbf{D}_{i,t-1},X_i)/\hat\rho_{d'_t}(\mathbf{D}_{i,t-1},X_i)}   \right].\notag
\end{align}
Furthermore, the algorithm can be easily modified to estimate $\Delta_{d_{t-s},d'_{t-s}, t}=E[Y_t(d_{t-s})-Y_t(d'_{t-s})|D_{t-s}=d_{t-s}]$, the ATET of a lagged treatment dose on a later outcome as defined in \eqref{ATElagged}, by using of the sample analogs of DR expressions \eqref{DiDidentDRcont3} or \eqref{DiDidentDRcont3modified} in the case of repeated cross-sections or panel data, respectively.

For the variances of our estimators, we adapt the asymptotic variance approximation suggested in \cite{zhang2025} to our settings with non-zero treatment doses in both the treatment and control groups. Considering the case of repeated cross-sections, we define
\begin{align}\label{scorerepeated}
&\psi^h(W,\Delta_{d_t,d_t',t}^h, \Pi_{d_t,t},\eta)\\
&=\frac{  \omega(D_t;d_t, h) \cdot I\{T=t\}\cdot [Y_T-\mu_{d_{t}}(t-1,\mathbf{D}_{t-1},X)-\mu_{d'_t}(t,\mathbf{D}_{t-1},X)+\mu_{d'_t}(t-1,\mathbf{D}_{t-1},X) ] }{ \Pi_{d_t,t}}\notag\\ 
&- \frac{\omega(D_t;d_t, h) \cdot I\{T=t-1\}\cdot \rho_{d_t,t}(\mathbf{D}_{t-1},X)\cdot \epsilon_{d_t,t-1,\mathbf{D}_{t-1},X}}{ \rho_{d_t,t-1}(\mathbf{D}_{t-1},X)\cdot \Pi_{d_t,t}}\\
&- \left(\frac{\omega(D_t;d'_t, h)\cdot  I\{T=t\}\cdot \rho_{d_t,t}(\mathbf{D}_{t-1},X)\cdot \epsilon_{d'_t,t,\mathbf{D}_{t-1},X}}{ \rho_{d'_t,t}(\mathbf{D}_{t-1},X) \cdot \Pi_{d_t,t}}\right. \notag\\
&- \left.\frac{\omega(D_t;d'_t, h)\cdot I\{T=t-1\}\cdot \rho_{d_t,t}(\mathbf{D}_{t-1},X)\cdot \epsilon_{d'_t,t-1,\mathbf{D}_{t-1},X}}{ \rho_{d_t',t-1}(\mathbf{D}_{t-1},X)\cdot \Pi_{d_t,t}}\right), 
\end{align}
with $\Delta_{d_t,d_t', t}^h= E\left[ \psi^h(W,\Delta_{d_t,d_t',t}^h, \Pi_{d_t,t},\eta) \right]$ 
being the smoothed ATET with bandwidth $h$. 
Then $\phi^h(W,\Delta_{d_t,d_t',t}^h, \Pi_{d_t,t},\eta):=\psi^h(W,\Delta_{d_t,d_t',t}^h, \Pi_{d_t,t},\eta)-\Delta_{d_t,d_t', t}^h$ is the smoothed Neyman-orthogonal score function for $\Delta_{d_t,d_t', t}^h$ with bandwidth $h$. The asymptotic variance, denoted by $\sigma_h^2$, of our DML estimator of $\Delta_{d_t,d_t', t}$ corresponds to 
\begin{align}\label{varianceformula}
    \sigma_h^2 = E\left[\left(\phi^h(W,\Delta_{d_t,d_t',t}^h, \Pi_{d_t,t},\eta) - \frac{\Delta_{d_t, d_t', t}^h}{\Pi_{d_t,t}}\left(\omega(D;d,h)\cdot I\{T=t\} - E[\omega(D;d,h)\cdot I\{T=t\}]\right)\right)^2 \right].
\end{align}
It is worth noting that the second term in the asymptotic variance in equation \eqref{varianceformula}
\[
\frac{\Delta_{d_t, d_t', t}^h}{\Pi_{d_t,t}}\left(\omega(D;d,h)\cdot I\{T=t\} - E[\omega(D;d,h)\cdot I\{T=t\}]\right)
\]
arises from a Taylor expansion of the score w.r.t.\ the (low dimensional) density $P(D_t=d_t, T=t)$. 

A cross-fitted variance estimator can be constructed based on the sample analog of equation \eqref{varianceformula}:
\begin{align}\label{varestimator}
    \hat{\sigma}_h^2 = \frac{1}{K}\sum_{k=1}^K \sum_{i\in I_k}\left[\left(\phi^h(W_i,\hat{\Delta}_{d_t,d_t',t}^h, \hat{\Pi}_{k,d_t,t},\hat{\eta}_k) -  \frac{\hat{\Delta}_{d_t, d_t', t}^h}{\hat{\Pi}_{k,d_t,t}}\left(\omega(D;d,h)\cdot I\{T=t\} - \hat{\Pi}_{k,d_t,t}\right)\right)^2 \right],
\end{align}
 where $\hat\Delta_{d_t,d_t', t}^h$ is the cross-fitted ATET estimate (see step 4 of the algorithm). The term $\phi^h(W_i,\hat{\Delta}_{d_t,d_t',t}^h, \hat{\Pi}_{k,d_t,t},\hat{\eta}_k)$ denotes the estimate of score function $\phi^h(W,\Delta_{d_t,d_t',t}^h, \Pi_{d_t,t},\eta)$ for subject $i$, which is computed using the subsample estimates of the nuisance parameters $\hat{\eta}_k$ (see step 3 of the algorithm). Similarly $\hat{\Pi}_{k,d_t,t} =|\mathcal{W}_k^{C}|^{-1}\sum_{i} \omega(D_i;d,h)\cdot I\{T_i=t\}$ is the subsample estimate of the density $\Pi_{d_t,t}$. In \cite{zhang2025}, such types of variance estimators are shown to be consistent, provided that the kernel bandwidth $h$ satisfies $h^{-2}\varepsilon_n^2 + h^{-3}n^{-1} = o(1)$ for $\varepsilon_n = o(n^{-1/4})$.

In the case of panel data, the results in \cite{zhang2025} imply that the asymptotic variance of the ATET estimator of $\Delta_{d_t,d_t', t}$ takes the following form:
\begin{align}\label{varpanel}
        \sigma_h^2 = E\left[\left(\phi^h(W,\Delta_{d_t,d_t',t}^h, P_{d_t},\eta) -  \frac{\Delta_{d_t, d_t', t}^h}{P_{d_t}}\left(\omega(D_t;d_t,h) - E[\omega(D_t;d_t,h)]\right)\right)^2 \right],
\end{align}
where $\Delta_{d_t,d_t',t}^h=E[\psi^h(W,\Delta_{d_t,d_t',t}^h, P_{d_t},\eta)]$ is the smoothed ATET under bandwidth $h$ and $\phi^h(W,\Delta_{d_t,d_t',t}^h, P_{d_t},\eta)=\psi^h(W,\Delta_{d_t,d_t',t}^h, P_{d_t},\eta)-\Delta_{d_t,d_t',t}^h$ is the score function, with 
\begin{align}\label{scorepanel}
  \psi^h(W,\Delta_{d_t,d_t',t}^h, P_{d_t},\eta)  & = \frac{  \omega(D_t;d_t, h) \cdot [Y_t-Y_{t-1}-m_{d_{t}'}(t,\mathbf{D}_{t-1},X)] }{ P_{d_t}}\\ 
&- \frac{\omega(D_t;d_t', h)\cdot  p_{d_t}(\mathbf{D}_{t-1},X)\cdot [Y_t-Y_{t-1}-m_{d_{t}'}(t,\mathbf{D}_{t-1},X)] }{p_{d_t'}(\mathbf{D}_{t-1},X) \cdot P_{d_t}}.\notag
\end{align}
A cross-fitted estimator of the variance in equation \eqref{varpanel} can be constructed analogously as outlined for the case of repeated cross-sections. 

Lastly, the corresponding estimators and variance expressions for the ATET with lagged treatment $\Delta_{d_{t-s},d'_{t-s}, t}=E[Y_t(d_{t-s})-Y_t(d'_{t-s})|D_{t-s}=d_{t-s}]$ can be easily obtained by substituting the expressions shown earlier with the lagged versions.

As an alternative approach to the asymptotic variance approximation, one can employ a multiplier bootstrap procedure for constructing confidence intervals. Let $\{\xi_i\}_{i=1}^n$ be an i.i.d.\ sequence of sub-exponential random variables independent from the data $\{W_i\}_{i=1}^n$, such that $E[\xi] = Var(\xi) = 1$. In each bootstrap replication $b = 1, \cdots, B$, one draws such a sequence $\{\xi_i\}_{i=1}^n$ and then constructs a bootstrap-specific ATET estimators, denoted by $\hat{\Delta}_{d_t, d_t', t}^{b,h}$, using the following expression:
\begin{align}\label{est:bootstrap}
    \hat{\Delta}_{d_t, d_t', t}^{b,h} = \frac{1}{K}\sum_{k=1}^K\sum_{i\in\mathcal{W}_k^{C}} \xi_i \cdot \psi^h(W_i,\hat{\Delta}_{d_t,d_t',t}^h, \hat{\Pi}_{k,d_t,t},\hat{\eta}). 
\end{align}
Let $\hat{c}_{\alpha}$ denote the $\alpha$-th quantile of the difference $\{\hat{\Delta}_{d_t, d_t', t}^{b,h} - \hat{\Delta}_{d_t, d_t', t}^h\}_{b=1}^B$, a $1-\alpha$ confidence interval may be constructed as $[\hat{\Delta}_{d_t, d_t', t}^h - \hat{c}_{1-\alpha/2}, \hat{\Delta}_{d_t, d_t', t}^h - \hat{c}_{\alpha/2}]$. 

Moreover, in the continuous treatment setting, it is also useful to consider uniform inference, and the bootstrap estimator defined in (\ref{est:bootstrap}) can be used to construct the uniform confidence bands. For simplicity, we focus on the two-period repeated cross-sectional case comparing the treated group with intensity $d$ and the control group with intensity $0$. In particular, this extends the repeated cross-sectional results in \cite{zhang2025} to allow time-varying covariates using efficient scores.

Recall that our causal parameter in this setting is defined as
\begin{align}
    \Delta_{d_t, 0, t} = E[Y_t(d_t) - Y_t(0)|D_t = d_t, T=t],
\end{align}
and we aim to establish valid uniform confidence bands for $\Delta_{d_t, 0, t}$ over the support of $D_t$. Let $\hat{\Delta}_{d_t, 0, t}^h$, $\hat{\Delta}_{d_t, 0, t}^{b,h}$, and $\hat{\sigma}_h^2(d_t,0,t)$ denote our DML estimator, the multiplier bootstrap estimator, and the cross-fitted variance estimator respectively. We consider the following procedure that follows \cite{CCK2014a, CCK2014b}, \cite{FHLZ22}, and \cite{zhang2025} to establish valid uniform confidence bands.
\begin{itemize}
    \item[1.] Estimate $\hat{\Delta}_{d_t, 0, t}^h$ and $\hat{\sigma}_h^2(d_t,0,t)$ on a finite grid of values $d_t\in \bar{\mathcal{D}}_t \subset supp(D_t)$.
    \item[2.] For each $b=1, \cdots, B$, draw an i.i.d. sequence of multipliers $\{\xi\}_{i=1}^N$ from a $N(1,1)$ distribution, and construct $\hat{\Delta}_{d_t, 0, t}^{b,h}$ for all $d_t\in \bar{\mathcal{D}}_t$.
    \item[3.] Compute $\hat{c}(1-\alpha)$, which we denote as the $1-\alpha$-quantile of 
    \[
    \Bigg\{\max_{d_t\in \bar{\mathcal{D}}_t} \frac{\sqrt{N}|\hat{\Delta}_{d_t, 0, t}^h - \hat{\Delta}_{d_t, 0, t}^{b,h}|}{\hat{\sigma}_h^2(d_t,0,t)} \Bigg \}_{b=1}^B.
    \]
    \item[4.] For all $d_t\in supp(D_t)$, construct the $1-\alpha$ uniform confidence band as
    \[
    [\hat{\Delta}_{d_t, 0, t}^h - \hat{c}(1-\alpha)\hat{\sigma}_h^2(d_t,0,t)/\sqrt{N}, \quad \hat{\Delta}_{d_t, 0, t}^h + \hat{c}(1-\alpha)\hat{\sigma}_h^2(d_t,0,t)/\sqrt{N}].
    \]
\end{itemize}
We omit the formal theoretical discussion here and point the readers to \cite{zhang2025}, which establishes valid uniform asymptotic linear expansion of the bootstrap estimators analogous to ours, and the validity of the uniform confidence bands constructed above follows from the results in \cite{CCK2014a,CCK2014b} and \cite{FHLZ22}.

\section{Asymptotic Theory}\label{asymp}
This section establishes the asymptotic properties of our estimators. The results largely follow \cite{zhang2025} and we sketch the main idea of the proofs in Appendix \ref{asymproof}. First, we impose a set of regularity conditions on the kernel function and the boundedness and smoothness of the model parameters. These conditions are used to establish three main results. First, the bias of ATET from using a kernel approximation becomes negligible asymptotically if the kernel bandwidth is undersmoothed. Second, the ATET estimators are asymptotically normal. Third, the variance estimators based on the asymptotic variance formulas proposed  in  Section \ref{sec:meth} are consistent. We focus on the case without time lags, and the results for the case with time lags follow immediately by appropriately defining treatment and outcome periods according to the time lag $t-s$. Recall that in the previous sections, we define the weighting function as $\omega(D;d,h)=\frac{1}{h}\mathcal{K}\left(\frac{D-d}{h}\right)$ with $h$ being the bandwidth. Using this weighting function, we consider the bandwidth-dependent approximations of the generalized propensitiy scores, denoted by $\rho_{d,t}^h(\mathbf{D}_{t-1},X)= E[\omega(D;d,h) \cdot I\{T=t\}|\mathbf{D}_{t-1},X]$ and $p_{d}^h(\mathbf{D}_{t-1},X)= E[\omega(D;d,h)|\mathbf{D}_{t-1},X]$, for the cross-sectional and panel settings respectively. In addition, we use $\hat\rho_{d,t}^h(\mathbf{D}_{t-1},X)$ and $\hat p_{d}^h(\mathbf{D}_{t-1},X)$ to denote the estimates of $\rho_{d,t}^h(\mathbf{D}_{t-1},X)$ and $p_{d}^h(\mathbf{D}_{t-1},X)$.

\begin{assumption} {\bf (Kernel):}\label{ass9}\\
The kernel function $\mathcal{K}(\cdot)$ satisfies: (a) $\mathcal{K}(\cdot)$ is bounded and differentiable; (b) $\int \mathcal{K}(u) du = 1$, $\int u\mathcal{K}(u)du = 0$, $0<\int u^2 \mathcal{K}(u) du <\infty$.
\end{assumption} 

\begin{assumption} {\bf (Bounds and smoothness, repeated cross-sections):} \label{ass10}\\
(a) There exists $0<c<1$ and $0<C<\infty$ such that $\Pi_{d_t,t}>c$, $c<\rho_{d_t, t-1}^h(\mathbf{D}_{t-1},X)<C$, $\rho_{d_t', t}^h(\mathbf{D}_{t-1},X)>c$, $\rho_{d_t', t-1}^h(\mathbf{D}_{t-1},X)>c$, $|Y_T| < C$, $|\mu_{d_t}(t-1,\mathbf{D}_{t-1},X)|<C$, $|\mu_{d_t'}(t,\mathbf{D}_{t-1},X)|<C$, and $|\mu_{d_t'}(t-1,\mathbf{D}_{t-1},X)|<C$ almost surely; (b) $\Pi_{d_t,t} \in C^2(\text{supp}(D_t))$ and $|\Pi_{d_t,t}^{(2)}| < \infty$; (c) $\rho_{d_t, t-1}(\mathbf{D}_{t-1},d) \in C^2(\text{supp}(D_t))$ for all $(\mathbf{D}_{t-1}, x)$, and $\sup_{d_t, d_{t-1}, x} |\rho_{d_t, t-1}^{(2)}(\mathbf{D}_{t-1},x)| < \infty$; (d) joint density
$f_{Y_T, D_t}(y,d_t) \in C^2(\text{supp}(Y_T))$ and $\sup_{y, d_t \in \text{supp}(Y_T, D_t)} |f_{Y_T, D_t}^{(2)}(y,d_t)| < \infty$.
\end{assumption}

\begin{assumption}{\bf (Rates, repeated cross-sections):}\label{ass11}\\ 
(a) The kernel bandwidth $h = h_N\to 0$ satisfies $Nh\to\infty$ and $\sqrt{Nh^5} = o(1)$; (b) there exists a sequence $\varepsilon_n\to 0$, such that $h^{-1}\varepsilon_n^2 = o(1)$; (c) with probability tending to $1$, $\|\hat{\rho}_{d_t, t}^h(\mathbf{D}_{t-1},X) - \rho_{d_t, t}^h(\mathbf{D}_{t-1},X)\|_{P,2}\leq h^{-1/2}\varepsilon_n$, $\|\hat{\rho}_{d_t, t-1}^h(\mathbf{D}_{t-1},X) - \rho_{d_t, t-1}^h(\mathbf{D}_{t-1},X)\|_{P,2}\leq \varepsilon_n$, $\|\hat{\rho}_{d_t', t-1}^h(\mathbf{D}_{t-1},X) - \rho_{d_t', t-1}^h(\mathbf{D}_{t-1},X)\|_{P,2}\leq \varepsilon_n$, $\|\hat{\mu}_{d_t}(t-1,\mathbf{D}_{t-1},X) - \mu_{d_t}(t-1,\mathbf{D}_{t-1},X)\|_{P,2}\leq \varepsilon_n$, $\|\hat{\mu}_{d_t'}(t-1,\mathbf{D}_{t-1},X) - \mu_{d_t'}(t-1,\mathbf{D}_{t-1},X)\|_{P,2}\leq \varepsilon_n$, $\|\hat{\mu}_{d_t'}(t,\mathbf{D}_{t-1},X) - \mu_{d_t'}(t,\mathbf{D}_{t-1},X)\|_{P,2}\leq \varepsilon_n$; (d) with probability tending to $1$, $ c< \hat{\rho}_{d_t, t-1}^h(\mathbf{D}_{t-1},X) <C$, $ \hat{\rho}_{d_t', t}^h(\mathbf{D}_{t-1},X) >c$, and $\hat{\rho}_{d_t', t-1}^h(\mathbf{D}_{t-1},X) >c$ almost surely; (e) with probability tending to 1, $\|\hat{\mu}_{d_t}(t-1,\mathbf{D}_{t-1},X)\|_{P,\infty}<C$, $\|\hat{\mu}_{d_t'}(t-1,\mathbf{D}_{t-1},X)\|_{P,\infty}<C$, and $\|\hat{\mu}_{d_t'}(t,\mathbf{D}_{t-1},X)\|_{P,\infty}<C$.
\end{assumption}

\begin{assumption}{\bf (Bounds and smoothness, panel):} \label{ass12}\\ (a) There exist constants $0<c<1$ and $0<C<\infty$, $P_{d_t}>c$, $|Y_{t-1}| < C$, $|Y_{t}| < C$, $p_{d_t'}^h(\mathbf{D}_{t-1},X)> c$, $|p_{d_t}^h(\mathbf{D}_{t-1},X)|<C$, and $|m_{d_t'}(t,\mathbf{D}_{t-1}, X)|<C$ almost surely; (b) $P_{d_t} \in C^2(\text{supp}(D_t))$ and $|P_{d_t}^{(2)}| < \infty$; (c) $p_{d_t}(\mathbf{D}_{t-1},x) \in C^2(\text{supp}(D_t))$ for all $(\mathbf{D}_{t-1},x)$ and $\sup_{\text{supp}(\mathbf{D}_{t-1}, X)} |p_{d_t}^{(2)}(\mathbf{D}_{t-1},X)| < \infty$; (d) for $\Delta Y = Y_t - Y_{t-1}$, the joint density
$f_{\Delta Y, D_t}(s, d_t) \in C^2(\text{supp}(\Delta Y))$ and $\sup_{s,d_t \in \text{supp}(\Delta Y, D_t)} |f_{\Delta Y, D_t}^{(2)}(s, d_t)| < \infty$.
\end{assumption}

\begin{assumption}{\bf (Rates, panel):} \label{ass13}\\ (a) The kernel bandwidth $h = h_N\to 0$ satisfies $Nh\to\infty$ and $\sqrt{Nh^5} = o(1)$; (b) there exists a sequence $\varepsilon_n\to 0$ such that $h^{-1}\varepsilon_n^2 = o(1)$; (c) with probability tending to $1$, $\|\hat{p}_{d_t}^h(\mathbf{D}_{t-1},X) - p_{d_t}^h(\mathbf{D}_{t-1},X)\|_{P,2}\leq h^{-1/2}\varepsilon_n$, $\|\hat{p}_{d_t'}^h(\mathbf{D}_{t-1},X) - p_{d_t'}^h(\mathbf{D}_{t-1},X)\|_{P,2}\leq \varepsilon_n$, $\|\hat{m}_{d_t'}(t,\mathbf{D}_{t-1}, X) - m_{d_t'}(t,\mathbf{D}_{t-1}, X)\|_{P,2}\leq \varepsilon_n$; (d) with probability tending to $1$, $\sup_{\text{supp}(\mathbf{D}_{t-1}, X)} \hat{p}_{d_t'}^h(\mathbf{D}_{t-1},X) > c$, $\|\hat{p}_{d_t}^h(\mathbf{D}_{t-1},X)\|_{P,\infty}<C$, $\|\hat{m}_{d_t'}(t,\mathbf{D}_{t-1}, X)\|_{P,\infty}<C$.
\end{assumption}

Assumption \ref{ass9} on the kernel function is standard and many commonly used smoothing functions satisfy these conditions, e.g., the Gaussian or Epanechnikov kernels. Moreover, we impose two sets of regularity conditions in both repeated cross-sections and panel settings. First, we require that bandwidth-dependent nuisance parameters in the population are bounded and satisfy certain smoothness conditions, see Assumptions \ref{ass10} and \ref{ass12}. Second, we impose high-level assumptions on the quality of the estimators of the nuisance parameters, permitting those estimators to converge at a slower rate compared to a setting without Neyman orthogonality, see Assumptions \ref{ass11} and \ref{ass13}. A first result obtained from these conditions is that the biases in repeated cross-sections and panel settings are of order $O(h^2)$, such that they vanish asymptotically when using an undersmoothed kernel bandwidth for ATET estimation.

\begin{lemma}\label{lemma:bias}
Suppose Assumptions \ref{ass9}, \ref{ass10}, \ref{ass11} hold for the repeated cross-sections case, and Assumptions \ref{ass9}, \ref{ass12}, \ref{ass13} hold for the panel case. Then, the bias satisfies $B(h) = \Delta_{d_t,d_t',t} - \Delta_{d_t,d_t',t}^h = O(h^2)$ for $d_t, d_t'\in \text{supp}(D_t)$.
\end{lemma}

The next two theorems are the main results of this section. The first theorem states that our proposed ATET estimators are asymptotically normal and converge at $\sqrt{nh}$ rate to the true ATET, while the second theorem states that our cross-fitted variance estimators proposed in the previous section are consistent.

\begin{theorem}\label{thm:normality}
Suppose Assumptions \ref{ass4}, \ref{ass5}, \ref{ass6}, \ref{ass7} hold. Moreover, suppose Assumptions \ref{ass9}, \ref{ass10}, \ref{ass11} hold for the repeated cross-sections case, and Assumptions \ref{ass9}, \ref{ass12}, \ref{ass13} hold for the panel case. Then for $\varepsilon_n = o(n^{-1/4})$, the ATET estimators defined in Section \ref{sec:meth} satisfy
\begin{align}
    \frac{\hat{\Delta}_{d_t,d_t',t} - \Delta_{d_t,d_t',t}}{\sigma_h/\sqrt{n}} \quad \rightarrow^d \quad N(0,1)
\end{align}
where $\sigma_h$ is defined as in (\ref{varianceformula}) and (\ref{varpanel}) for the repeated cross-sections case and panel case respectively. 
\end{theorem}

\begin{theorem}\label{thm:var}
Assume the conditions in Theorem \ref{thm:normality} hold and assume that $h^{-2}\varepsilon_n^2 + h^{-3}n^{-1} = o(1)$. Then the cross-fitted variance estimators defined in Section \ref{sec:meth} are consistent for the asymptotic variance $\sigma_h$ defined in (\ref{varianceformula}) and (\ref{varpanel}) for the repeated cross-sections case and panel case respectively.
\end{theorem}
\noindent A sketch of the proofs is provided in Appendix \ref{asymproof}.

\section{Simulation study}\label{sim}

This section provides a simulation study to investigate the finite sample behavior of our DiD estimators for repeated cross-sections and panel data. Starting with the case of repeated cross-sections, we consider the following data generating process (DGP):
\begin{eqnarray*}
X_j &=& 0.5T+Q_j\textrm{ for }j \in \{1,...,p\},\quad X=(X_1,...,X_p),\\
D &=& X\beta+0.5\cdot U + V ,\\
Y_T &=& X\beta+ (1+D^2)\cdot T + U+W,\\
Q_j,U,V,W &\sim& Unif(0,2)\textrm{ independent of each other},\\
T&\sim& binomial (0.5).
\end{eqnarray*}

The vector $X$ consists of time varying covariates $X_j$ with $j \in \{1,...,p\}$, where $p$ is the dimension of $X$. Each covariate $X_j$ depends on the time index $T$, which is either 0 for the pre-treatment and 1 for the post-treatment period (each with a probability of 50\%), as well as the random error $Q_j$, which is independently uniformly distributed with its support ranging from 0 to 2.

The continuous treatment $D$ is a function of covariates $X$ if the coefficient vector $\beta $ is different from zero and two uniformly distributed error terms, namely the fixed effect $U$ and a random component $V$. Both the covariates $X$ and the fixed effect $U$ also affect the outcome $Y_T$ and are, therefore, confounders of the treatment-outcome relation that is tackled by our covariate-adjusted DiD approach. In addition, the outcome is affected by the period indicator $T$, which reflects a time trend, and a random uniformly distributed error term $W$. Lastly, the treatment has a nonlinear effect on the post-treatment outcome, which is modeled by the interaction of $D^2$ and $T$. 

In our simulations, we set the number of covariates to $p=100$ and the $j$th element in the coefficient vector $\beta$ to $0.4/i^2$ for $j \in \{1,...,p\}$, which implies a quadratic decay of covariate importance in terms of confounding. We consider two sample sizes of $n=2000, 4000$. DiD estimation is based on the sample analog of equation \eqref{DiDidentDRcont2}. In our simulation design, there are only two periods $t$ such that $T=t$ and $T=t-1$ in equation \eqref{DiDidentDRcont2} correspond to $T=1$ and $T=0$, respectively. Furthermore, there is no history of previous treatments in $T=0$ such that $\mathbf{D}_{t-1}=\emptyset$ in our simulations. However, from an econometric perspective, there is no difference between controlling for the treatment history $\mathbf{D}_{t-1}$ or covariates $X$, which are both time-varying sets of confounders. For this reason, not considering $\mathbf{D}_{t-1}$ in our simulations comes without loss of generality as we could easily relabel some elements in $X$ to represent $\mathbf{D}_{t-1}$.

For estimation, we apply the \textit{didcontDML} function from the \textit{causalweight} package by \citet{BodoryHuber2018} using the statistical software \textsf{R} (\cite{R2020}). We implement this DML-based DiD estimator for repeated cross-sections with two-fold cross-fitting and estimate the nuisance parameters via cross-validated lasso regression. We employ a second-order Epanechnikov kernel as the weighting function $\omega$ for smoothing over the continuous treatment, with a bandwidth given by $2.34 \cdot n^{-0.25}$. Choosing 2.34 as the bandwidth constant corresponds to a \cite{Silverman86}-type rule of thumb for second-order Epanechnikov kernels, as also applied, for instance, in \citet{huber2020direct}, while the rate $n^{-0.25}$ ensures undersmoothing to avoid asymptotic bias induced by kernel smoothing. In addition to estimation based on this rule, we consider a specification with stronger undersmoothing by multiplying the resulting bandwidth by 0.5 (which amounts to halving the constant in the bandwidth formula), yielding $2.34 \cdot n^{-0.25} \cdot 0.5$. 

Furthermore, when estimating the generalized propensity scores $\rho_{d,t}(X)$ via lasso, we consider both linear and log-linear specifications that assume normally distributed unobservables $V$ (an assumption that is violated here, as $V$ is uniformly distributed). The ATET estimator additionally imposes a trimming rule to safeguard against disproportionately influential observations in the inverse propensity weighting (IPW). Specifically, observations receiving a weight larger than 10\% in the IPW-based computation of any conditional mean outcome given the treatment state and time period are dropped. Such a trimming rule to ensure common support are also considered in \cite{HuLeWu10}. Finally, standard errors for the ATET estimates are obtained using the estimated asymptotic variance approximation given in equation~\eqref{varestimator} of Section~\ref{sec:meth}.

The upper panel of Table \ref{tab:sim} provides the simulation results for repeated cross-sections, when estimating the ATET $\Delta_{3,2,1}=E[Y_1(3)-Y_0(2)|D_1=3]=3^2-2^2=5$, based on linear lasso regression with weaker undersmoothing (`lasso'), loglinear lasso regression with weaker undersmoothing (`lnorm'), linear lasso regression with stronger undersmoothing (`under'), and loglinear lasso regression with stronger undersmoothing (`ln under'). The columns provide the estimation method (`method'), the bias of the estimator (`bias'), its standard deviation (`std'), its root mean squared error (`rmse'), and the average standard error (`avse'), respectively, for either sample size. 

For $n=2000$, we observe that the ATET estimators with weaker undersmoothing are non-negligibly biased, whereas those with stronger undersmoothing are less biased while having a slightly higher standard deviation. Overall, the strongly undersmoothed versions have a superior performance in terms of a lower root mean squared error (RMSE), which measures the overall estimation error as the square root of the sum of the squared bias and variance. Moreover, the ATET estimates are very similar when using normal or lognormal models for the estimation of the generalized propensity scores. Concerning statistical inference, the average standard error (`avse') is generally close to the actual standard deviation (`std'), suggesting that asymptotic variance estimation as outlined in Section \ref{sec:meth} performs well in our simulations. 
 When increasing the sample size to $n=8000$, the biases and standard deviations of all estimators decrease, but again, it is the strongly undersmoothed versions of the estimators that perform substantially better in terms of the RMSE than those with weaker undersmoothing. 

   \begin{table}[htbp]
		\begin{center}
			\caption{Simulation results}
            \vspace{5pt}
			\label{tab:sim}
			\begin{tabular}{c|cccc|cccc}
				\hline\hline
    &\multicolumn{8}{c}{repeated cross-sections}\\
        &\multicolumn{4}{c|}{$n=2000$}&\multicolumn{4}{c}{$n=8000$}\\
				method   & bias & std. dev. & rmse & avse & 
    bias & std & rmse &avse \\ 
				\hline
lasso & -0.406 & 0.227 & 0.465 & 0.222 & -0.202 & 0.152 & 0.253 & 0.151 \\ 
lnorm & -0.402 & 0.226 & 0.461 & 0.221 & -0.199 & 0.151 & 0.250 & 0.150 \\ 
under & -0.107 & 0.241 & 0.263 & 0.241 & -0.059 & 0.167 & 0.177 & 0.174 \\ 
ln under & -0.104 & 0.239 & 0.261 & 0.240 & -0.057 & 0.167 & 0.177 & 0.173  \\
 				\hline
     &\multicolumn{8}{c}{panel data}\\
        &\multicolumn{4}{c|}{$n=2000$}&\multicolumn{4}{c}{$n=8000$}\\
				method   & bias & std. dev. & rmse & avse &  
    bias & std & rmse &avse \\ 
				\hline
lasso  & -0.663 & 0.120 & 0.674 & 0.125 & -0.358 & 0.055 & 0.362 & 0.053\\ 
lnorm & -0.661 & 0.120 & 0.671 & 0.125 & -0.357 & 0.055 & 0.361 & 0.053\\ 
under & -0.177 & 0.100 & 0.203 & 0.106 & -0.094 & 0.056 & 0.110 & 0.056\\ 
ln under & -0.176 & 0.099 & 0.202 & 0.106 & -0.094 & 0.056 & 0.109 & 0.056  \\
     \hline
			\end{tabular}
		\end{center}
		\par
		  \begin{flushleft}
     \footnotesize \justifying \textit{Notes:} columns `method', `bias',  `std. dev.', `rmse', and `avse'  provide estimation method, the bias of the estimator, its standard deviation, its root mean squared error, and the average of the standard error, respectively. The `lasso' provides the results for lasso-based estimation when assuming a linear treatment model with normally distributed errors. The`lnorm' assumes instead a loglinear treatment model with normal errors. The `under' relies on a linear treatment model with strong undersmoothing, using half of the constant for bandwidth computation based on the rule of thumb for smoothing the continuous treatment. The `ln under' provides the results for a loglinear treatment model and strong undersmoothing. The true ATET is 5.
     \end{flushleft}
	\end{table}

Subsequently, we investigate the DiD estimator for panel data, considering a slightly modified DGP in which $X_j$ only affects the outcome in the second period, but not in the first period:
 \begin{eqnarray*}
X &=& (X_1,...,X_p),\\
D &=& X\beta+0.5\cdot U + V ,\\
Y_T &=& (1+D^2+X\beta)\cdot T + U+W,\\
X_j,U,V,W &\sim& Unif(0,2)\textrm{ independent of each other and for }j \in \{1,...,p\},\\
T &\in& \{0,1\}.
\end{eqnarray*}
We apply the \textit{didcontDMLpanel} function of the \textit{causalweight} package by \citet{BodoryHuber2018} and consider the same choices in terms of kernel functions, bandwidth selection, undersmoothing, generalized propensity score modeling, and trimming as for the repeated cross-sections case.

The lower panel of Table \ref{tab:sim} provides the simulation results for the panel case when estimating the ATET $\Delta_{3,2,1}=5$. The absolute magnitudes of the biases and the RMSEs of the ATET estimators with weaker undersmoothing are again considerably larger that those of the strongly undersmoothed versions for either sample size. Moreover, the average standard errors based on asymptotic variance approximation for the panel case as outlined in Section \ref{sec:meth} closely match the actual standard deviations of the respective estimators. 
For this reason, the simulation results for the panel and repeated cross-section cases are qualitatively similar.

\section{Empirical application}\label{appl}

In this section, we present an empirical application of our method in the context of the COVID-19 pandemic in Brazil—one of the countries most severely affected by the crisis. We estimate the causal effect of second-dose COVID-19 vaccination rates on mortality across Brazilian municipalities. Our data are compiled from multiple sources providing municipality-level health and socio-economic information, including the laboratory DB Molecular – Diagnósticos do Brasil, the Brazilian Institute of Geography and Statistics (IBGE), DataSUS (Ministry of Health, Brazil), and the monitoring of COVID-19 cases and deaths by \citet{CotaCovid19br2020}. After cleaning and harmonizing the data, we obtain a balanced panel of 213 municipalities observed daily from August~4,~2020 to March~23,~2022, covering 597 days and yielding 127{,}161 observations.

\begin{figure}
    \centering
    \includegraphics[width=0.8\linewidth]{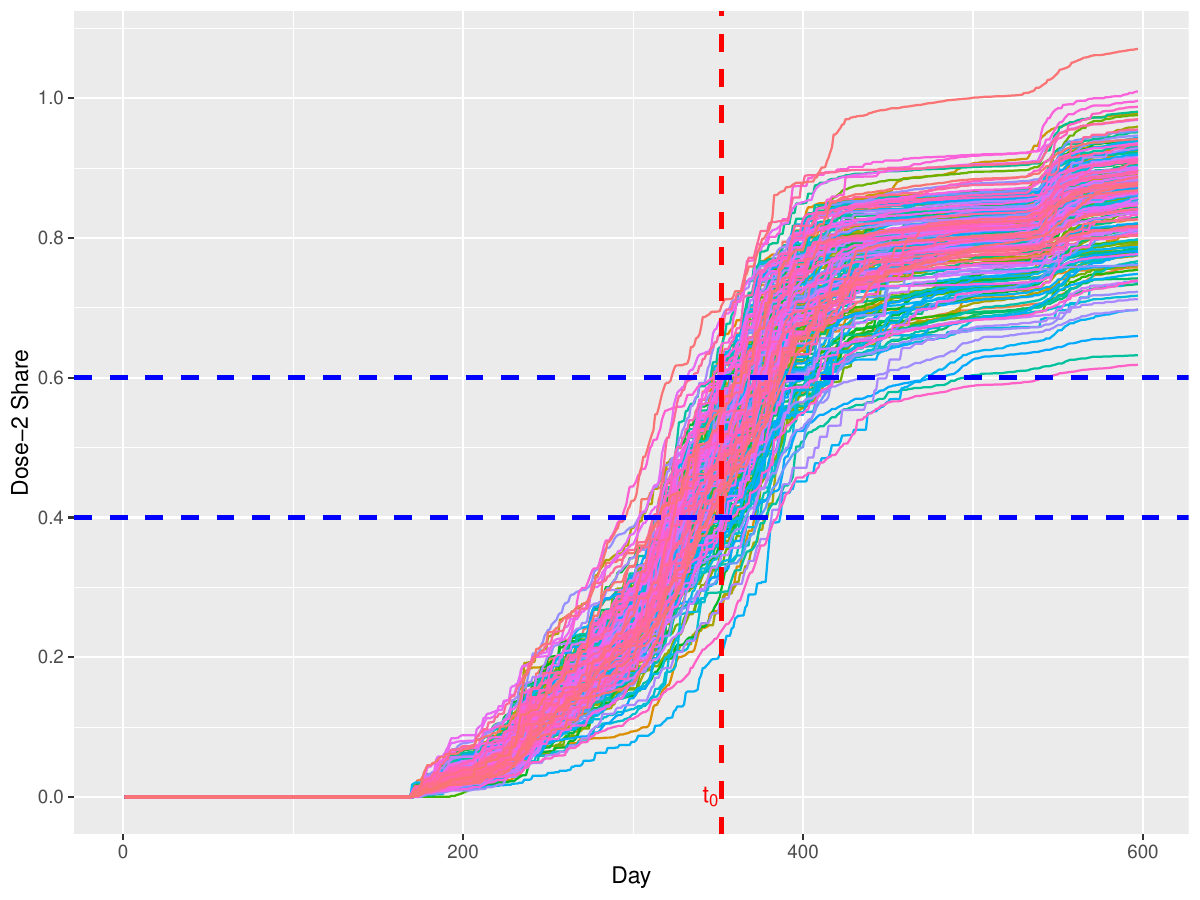}
    \caption{Dose-2 Vaccination Rate Over Time by Municipality}
    \label{fig:dose2}
\end{figure}

We define the continuous and time-varying treatment variable as the municipality-specific vaccination rate, measured as the ratio of cumulative dose-2 vaccinations administered in a municipality up to a given day to the municipality’s population size according to the 2020 census. Figure~\ref{fig:dose2} displays vaccination rates over time (across 597 days) during the sample period, highlighting considerable heterogeneity in the pace of vaccination across municipalities. Given the limited number of municipalities in our sample, we adopt a stacked design that permits pooling across multiple periods to increase statistical power. Specifically, we designate the first day on which any municipality reaches a $70\%$ (or 0.7) dose-2 vaccination rate as the reference day $t_0$, as indicated by the dashed vertical line in Figure~\ref{fig:dose2}. We then stack the data at seven-day intervals over five periods, $(t_0, t_7, t_{14}, t_{21}, t_{28})$, which increases the effective evaluation sample five-fold relative to a single day, yielding $5 \cdot 213 = 1065$ panel units. Furthermore, this design avoids common support issues at the treatment intensities under comparison, which are chosen to be 60\% (or 0.6) and 40\% (or 0.4), as indicated by the horizontal dashed lines in Figure~\ref{fig:dose2}.

Our outcome variable is defined as the number of new COVID-19-related deaths per $100{,}000$ population in a given municipality at specific post-treatment horizons—namely one, two, four, seven, ten, fifteen, thirty, and sixty days after treatment. This allows us to study the dynamic evolution of the treatment effect. We control for the following covariates in a data-adaptive manner, conditional on which the parallel trends assumption is required to hold: (1) 45 demographic and socioeconomic characteristics measured at the municipal level, such as fertility, education, income, and life expectancy; (2) a set of dummy variables indicating the stacked reference days; (3) treatment histories capturing dose-2 vaccination rates one, two, seven, and fourteen days prior to the treatment day; and (4) dose-1 vaccination rates on the treatment day as well as their corresponding histories (dose-1 vaccination rates one, two, seven, and fourteen days prior to the treatment day), to avoid confounding from dose-1 vaccination.

Table~\ref{tab:main_descriptive} reports descriptive statistics (mean, standard deviation, minimum, and maximum) for several key variables - including the outcome, the treatment, and a selected set of covariates reflecting the dose-1 vaccination  rate, socioeconomic conditions, and infrastructure - computed across the 213 municipalities and all 597 days of the balanced panel. While the mortality outcome and vaccination rates can vary at the daily level, municipalities' socioeconomic and infrastructure characteristics are measured only once in 2022 and are therefore time-invariant (although they may exert time-varying influences on outcomes, which is why they are considered as potential control variables). The table shows that municipalities represent diverse socioeconomic contexts, with substantial variation in per capita income and infrastructure (such as sewerage network coverage and access to piped water), which are relevant factors for healthcare capacity and pandemic management. See Appendix~\ref{stats_did} for a more comprehensive set of descriptive statistics covering a broader range of variables.

\begin{table}[h]
\centering
\caption{Descriptive statistics}
    \label{tab:main_descriptive}
      \begin{tabular}{l c c c c}
        \hline \hline
         variable  &  mean  & std. dev. & minimum & maximum \\
        \hline
        COVID-19 deaths / population      & 0.0020  & 0.0014  & 0.0000  & 0.0075  \\
        dose-1 vaccination  rate       & 0.26  & 0.30  & 0.00  & 1.30  \\
        dose-2 vaccination  rate       & 0.36  & 0.35  & 0.00  & 1.07  \\
        municipal HDI              & 0.74  & 0.05  & 0.57  & 0.86  \\
        Gini index                 & 0.50  & 0.05  & 0.35  & 0.63  \\
        per capita income          & 807.73   & 282.31    & 257.64   & 2043.74  \\
        proportion of poor (\%)    & 8.69  & 8.08  & 0.44  & 43.46 \\
        sewage network (\%)        & 68.49 & 28.91 & 0.71  & 99.83 \\
        piped water (\%)           & 95.63 & 5.56  & 60.18 & 99.99 \\
        \hline
    \end{tabular}
\end{table}

As in the simulation study in the previous section, we use the \textit{didcontDML} function from the \textit{causalweight} package by \citet{BodoryHuber2018} to estimate the causal effect of the vaccination rate, employing two-fold cross-fitting to mitigate overfitting. To flexibly adjust for high-dimensional control variables, the nuisance parameters are estimated nonparametrically using random forests as the machine learner (see, e.g., \citealp{Ho1995, Breiman2001}), assuming normally distributed unobservables in the generalized propensity score model. The kernel bandwidth used for smoothing over treatment intensity is set to $2.34 \cdot n^{-0.25} \cdot 0.7$, that is, the rule-of-thumb constant for bandwidth selection multiplied by $0.7$ to implement undersmoothing based on $n^{-0.25}$. This choice lies between the weaker and stronger undersmoothing approaches considered in the simulation study. Finally, we apply a trimming rule that drops observations receiving more than 10\% weight in the IPW-based estimation of the mean potential outcome under non-treatment.

Table~\ref{tab:deaths_06} reports the estimated ATETs (measured in COVID-19 deaths per $100{,}000$ population) based on a comparison of a higher dose-2 vaccination rate of 60\% ($d_{\text{treat}}=0.6$) versus a lower rate of 40\% ($d_{\text{control}}=0.4$) across post-treatment horizons. The corresponding standard errors (SE) and p-values, clustered at the municipality level, are also reported. The results indicate that higher dose-2 coverage has no statistically significant short-run effects from one to fifteen days after treatment. However, statistically significant negative effects emerge at longer horizons, after thirty days (at the 5\% level) and sixty days (at the 10\% level). Specifically, the DML estimates suggest a reduction of approximately 1.5 COVID-19 deaths per $100{,}000$ inhabitants in municipalities with a 60\% vaccination rate relative to a rate of 40\%, after one to two months. These findings are consistent with the expected time lag between infection and death documented in the COVID-19 literature; see, e.g., \citet{huber2020timing}.

\begin{table}[h]
\centering
\caption{Effects of dose-2 on COVID-19 deaths with $d_{treat}=0.6$, $d_{control}=0.4$}
\label{tab:deaths_06}
\begin{tabular}{l l  c c}
\hline\hline
{post-treatment days} &  {ATET} & {SE} & {$p$-value} \\
\hline
one day &  0.057 &	0.355 &	0.874 \\
two days &  2.657 &	2.469 &	0.282 \\
four days &  2.798 &	2.543 &	0.271 \\
seven days &  2.962 &	4.437 &	0.504 \\
ten days &  -0.072 &	2.733 &	0.979 \\
fifteen days &  -4.352 &	3.077 &	0.157 \\
thirty days &  -1.418 &	0.632 &	0.025 \\
sixty days &  -1.583 &	0.824 &	0.055 \\
\hline
\end{tabular}
\end{table}

\section{Conclusion}\label{sec:conc}




In this paper, we have developed a generalized difference-in-differences framework to estimate the causal effects of time-varying continuous treatments. Our approach focused on identifying the Average Treatment Effect on the Treated (ATET) by comparing distinct treatment intensities, relying on a conditional parallel trends assumption that robustly accounts for complex treatment histories and covariate evolution. By integrating double/debiased machine learning with kernel-based weighting, we provided a rigorous estimation strategy that remains consistent even with high-dimensional nuisance parameters. Finally, we illustrated our method by applying it to Brazilian municipality-level data to estimate the effect of second-dose COVID-19 vaccination rates on mortality. The results indicate that higher vaccination coverage does not significantly affect mortality in the very short term but leads to a statistically significant reduction in COVID-19 deaths after several weeks, consistent with the expected time lag between infection and fatal outcomes.


\pagebreak

\bibliographystyle{plainnat}
\bibliography{ref}

\pagebreak

{\large \renewcommand{\theequation}{A-\arabic{equation}}
\setcounter{equation}{0} \appendix }
\appendix \numberwithin{equation}{section}

\section{Appendix}

\subsection{Neyman orthogonality of the DR expressions}\label{Neymanrepeated}

We first show the Neyman orthogonality of the DR expressions \eqref{DiDidentDRcont}, \eqref{DiDidentDRcont2}, and \eqref{DiDidentDRcont3} for repeated cross-sections. We consider the ATET $\Delta_{d,t}=E[Y_t(d)-Y_t(0)|D=d]$ provided in equation \eqref{ATEsimple}, which is identified under Assumptions \ref{ass1} to \ref{ass4}. In the subsequent discussion, all bounds hold uniformly over all probability laws $P \in \mathcal{P}$, where $\mathcal{P}$ is the set of all possible probability laws - we omit $P$ for brevity. We define the nuisance parameters $$\eta=(\mu_{d'}(t',X),\rho_{d,t}(X),\rho_{d',t'}(X))$$ and denote the true values of the nuisance parameters in the population by $$\eta^0=(\mu^0_{d'}(t',X),\rho^0_{d,t}(X),\rho^0_{d',t'}(X)).$$ We also define $\Psi_{d',t'}=E[\mu_{d'}(t',X)|D=d,T=t]$ to be the average of the conditional mean $\mu_{d'}(t',X)$ in treatment group $D=d$ and time period $T=t$. We note that the ATET provided in equation \eqref{DiDidentpostreg} may be expressed as a function of $\Psi_{d',t'}$: 
\begin{align}\label{nuisancepar}
&\Delta_{d,t}=E[  \mu_d(t,X)-\mu_d(0,X) - (\mu_0(t,X)-\mu_0(0,X))|D=d, T=t ]\notag\\
&=E[Y_T|D=d,T=t]-\Psi_{d,0}- \Psi_{0,t} + \Psi_{0,0},
\end{align}
where the second equality follows from the law of iterated expectations. Therefore, considering Neyman-orthogonal functions for $\Psi_{d',t'}$ yields a Neyman-orthogonal expression of the ATET, which is an additive function of $\Psi_{d',t'}$ for $d'$ $\in$ $\{d,0\}$ and $t'$ $\in$ $\{t,0\}$. 

We denote by $\phi_{d',t'}$ the Neyman-orthogonal score function for $\Psi_{d',t'}$, which corresponds to the following expression for a continuous treatment $D$ when considering values $d$ and $d'$:    
\begin{align}\label{neymanscore}
\phi_{d',t'}&= \frac{  \omega(D;d, h) \cdot I\{T=t\}\cdot \mu_{d'}(t',X)}{\Pi_{d,t}}\notag\\
&+\frac{\omega(D;d', h) \cdot I\{T=t'\}\cdot \rho_{d,t}(X)\cdot (Y_T- \mu_{d'}(t',X))}{ \rho_{d',t'}(X)\cdot \Pi_{d,t}}-\Psi_{d',t'}.
\end{align}
We note that if $d$ and/or $d'$ are mass points (e.g., $d'=0$), then the kernel functions $\omega(D;d, h)$ and/or $\omega(D;d', h)$ should be replaced by indicator functions $I\{D=d\}$ and/or $I\{D=d'\}$. For the sake of simplicity, we omit this case in our discussion and focus on the continuous case. We also note that DR expression \eqref{DiDidentDRcont} is obtained by identifying $\Psi_{d',m'}$ as the solution to the moment condition $E[\phi_{d',m'}]=0$, and plugging it into equation \eqref{nuisancepar}. 

The Gateaux derivative of $E[\phi_{d',t'}]$ in the direction of $[\eta^0-\eta]$ is given by:
\begin{align}\label{derivscore}
&\partial E[\phi_{d',t'}] [\eta-\eta^0] = \lim_{h \rightarrow 0} E \left[  \frac{  \omega(D;d, h) \cdot I\{T=t\}\cdot [\mu_{d'}(t',X)-\mu^0_{d'}(t',X)]}{ \Pi_{d,t}}\right]\notag\\
&-\lim_{h \rightarrow 0} E \left[  \frac{  \omega(D;d', h) \cdot I\{T=t'\}\cdot\rho^0_{d,t}(X)\cdot [\mu_{d'}(t',X)-\mu^0_{d'}(t',X)]}{ \rho^0_{d',t'}(X) \cdot \Pi_{d,t}}\right] \notag\\
&+\lim_{h \rightarrow 0} E \left[  \frac{  \omega(D;d', h) \cdot I\{T=t'\}\cdot [Y_T-\mu^0_{d'}(t',X)]}{ \rho^0_{d',t'}(X) } \cdot \frac{[ \rho_{d,t}(X) - \rho^0_{d,t}(X)] } {\Pi_{d,t}}\right] \\
&-\lim_{h \rightarrow 0} E \left[  \frac{  \omega(D;d', h) \cdot I\{T=t'\}\cdot\rho^0_{d,t}(X)\cdot [Y_T-\mu^0_{d'}(t',X)]}{ \rho^0_{d',t'}(X) \cdot \Pi_{d,t} } \cdot \frac{ \rho_{d',t'}(X) -  \rho^0_{d',t'}(X)} { \rho^0_{d',t'}(X)  }\right]=0.\notag
\end{align}

Applying the law of iterated expectations to the term in the second line of equation \eqref{derivscore} yields:
\begin{align}
&-\lim_{h \rightarrow 0} E \left[  E\left[\frac{  \omega(D;d', h) \cdot I\{T=t'\}\cdot\rho^0_{d,t}(X)\cdot [\mu_{d'}(t',X)-\mu^0_{d'}(t',X)]}{ \rho^0_{d',t'}(X) \cdot \Pi_{d,t}}\ \bigg| X\right]\right]\notag\\
&=-\lim_{h \rightarrow 0} E \left[ \frac{ E[  \omega(D;d', h) \cdot I\{T=t'\}|X]\cdot\rho^0_{d,t}(X)\cdot [\mu_{d'}(t',X)-\mu^0_{d'}(t',X)]}{ \rho^0_{d',t'}(X) \cdot \Pi_{d,t}}\right]\notag\\
&=- E \left[  \frac{\rho^0_{d',t'}(X)\cdot\rho^0_{d,t}(X)\cdot [\mu_{d'}(t',X)-\mu^0_{d'}(t',X)]}{ \rho^0_{d',t'}(X) \cdot \Pi_{d,t}}\right]\notag\\
&=- E \left[  \frac{\rho^0_{d,t}(X)\cdot [\mu_{d'}(t',X)-\mu^0_{d'}(t',X)]}{ \Pi_{d,t}}\right].\notag
\end{align}
Applying the law of iterated expectations to the term on the left hand side of the first line yields:
\begin{align}
 &\lim_{h \rightarrow 0} E \left[  \frac{  E[\omega(D;d, h) \cdot I\{T=t\}|X]\cdot [\mu_{d'}(t',X)-\mu^0_{d'}(t',X)]}{ \Pi_{d,t}}\right]\notag\\
 = &  E \left[  \frac{\rho^0_{d,t}(X) \cdot [\mu_{d'}(t',X)-\mu^0_{d'}(t',X)]}{ \Pi_{d,t}}\right].\notag
\end{align}
Consequently, the terms in the first and in second line of equation \eqref{derivscore} cancel out. Considering the third line, we have that:
\begin{align}
&\lim_{h \rightarrow 0} E \left[  E \left[\frac{  \omega(D;d', h) \cdot I\{T=t'\}\cdot [Y_T-\mu^0_{d'}(t',X)]}{ \rho^0_{d',t'}(X) } \cdot \frac{[ \rho_{d,t}(X) - \rho^0_{d,t}(X)] } {\Pi_{d,t}}\bigg| X\right] \right] \notag\\
=&\lim_{h \rightarrow 0} E \left[  \frac{ E [ \omega(D;d', h) \cdot I\{T=t'\}\cdot [Y_T-\mu^0_{d'}(t',X)]|X]}{ \rho^0_{d',t'}(X) } \cdot \frac{[ \rho_{d,t}(X) - \rho^0_{d,t}(X)] } {\Pi_{d,t}} \right]. \notag
\end{align}
Concerning the first factor, we have by basic probability theory that:
\begin{align}
&\frac{\lim_{h \rightarrow 0} E [ \omega(D;d', h) \cdot I\{T=t'\}\cdot Y_T|X]}{ \rho^0_{d',t'}(X) }= E [ Y_T|D=d', T=t',X]=\mu^0_{d'}(t',X).\notag
\end{align}
Furthermore,
\begin{align}
&\frac{\lim_{h \rightarrow 0} E [ \omega(D;d', h) \cdot I\{T=t'\}\cdot \mu^0_{d'}(t',X)|X]}{ \rho^0_{d',t'}(X) }\notag\\
=& \frac{\lim_{h \rightarrow 0} E [ \omega(D;d', h) \cdot I\{T=t'\}|X]\cdot \mu^0_{d'}(t',X)}{ \rho^0_{d',t'}(X) }= \frac{\rho^0_{d',t'}(X) \cdot \mu^0_{d'}(t',X)}{ \rho^0_{d',t'}(X) }=\mu^0_{d'}(t',X).\notag
\end{align}
This implies that the first factor in the third line of equation \eqref{derivscore} is zero, such that the entire expression in the third line is also zero. In analogous manner, we show for the fourth line that:
\begin{align}
&-\lim_{h \rightarrow 0} E \left[  E \left[\frac{  \omega(D;d', h) \cdot I\{T=t'\}\cdot\rho^0_{d,t}(X)\cdot [Y_T-\mu^0_{d'}(t',X)]}{ \rho^0_{d',t'}(X) \cdot \Pi_{d,t} } \cdot \frac{ \rho_{d',t'}(X) -  \rho^0_{d',t'}(X)} { \rho^0_{d',t'}(X)  }\bigg| X\right]\right]\notag\\
&=- E \left[  \frac{  \rho^0_{d,t}(X)\cdot [\mu^0_{d'}(t',X)-\mu^0_{d'}(t',X)]}{   \Pi_{d,t} } \cdot \frac{ \rho_{d',t'}(X) -  \rho^0_{d',t'}(X)} { \rho^0_{d',t'}(X)  } \right].\notag
\end{align}
As the first factor is zero due to the fact that $[\mu^0_{d'}(t',X)-\mu^0_{d'}(t',X)]=0$, the expression in the fourth line of equation \eqref{derivscore} is also zero. Therefore, the Gateaux derivative of $E[\phi_{d',t'}]$ in the direction of $[\eta^0-\eta]$ is zero. Moreover, we trivially have that the Gateaux derivative of $E[Y_T|D=d,T=t]$ in the direction of $[\eta^0-\eta]$ is zero. We note that the DR expression \eqref{DiDidentDRcont} is equivalent to expression \eqref{nuisancepar} when using the Neyman-orthogonal score functions \eqref{neymanscore} for computing $\Psi_{d,0}$, $\Psi_{0,t}$ $\Psi_{0,0}$, noticing that $E[Y_T|D=d,T=t]=\lim_{h \rightarrow 0} E \left[\frac{  \omega(D;d, h) \cdot I\{T=t\}\cdot Y_T}{\Pi_{d,t}}\right]$. For this reason, the DR expression \eqref{DiDidentDRcont} satisfies the Neyman orthogonality.  

Next, consider the identification of the ATET $\Delta_{d_t,d_t', t}=E[Y_t(d_t)-Y_t(d_t')|D_t=d_t]$ in equation \eqref{ATEsameperiod} based on Assumptions \ref{ass4} to \ref{ass7} using DR expression \eqref{DiDidentDRcont2}. The steps for showing Neyman orthogonality are equivalent when replacing the nuisance parameters in equation \eqref{nuisancepar} by the following nuisance parameters, which additionally includes the treatment history $\mathbf{D}_{t-1}$ in the conditioning set:
\begin{align} 
  \eta =&  \{ \mu_{d_{t}}(t-1,\mathbf{D}_{t-1},X),\mu_{d'_t}(t,\mathbf{D}_{t-1},X),\mu_{d'_t}(t-1,\mathbf{D}_{t-1},X),\notag\\ 
  &\rho_{d_t,t}(\mathbf{D}_{t-1},X), \rho_{d_t,t-1}(\mathbf{D}_{t-1},X),\rho_{d'_t,t}(\mathbf{D}_{t-1},X),\hat\rho_{d'_t,t-1}(\mathbf{D}_{t-1},X)\}.\notag 
  \end{align}
 In addition, consider the identification of the ATET $\Delta_{d_{t-s},d'_{t-s}, t}=E[Y_t(d_{t-s})-Y_t(d'_{t-s})|D_{t-s}=d_{t-s}]$ in equation \eqref{ATElagged} based on Assumptions \ref{ass4}, \ref{ass6}, \ref{ass7}, and \ref{ass8} using DR expression \eqref{DiDidentDRcont3}. Also in this case, the steps for showing Neyman orthogonality are analogous, with the only difference to the previous setup being the introduction of a time lag between the treatment and post-treatment outcomes. 

Next, we demonstrate the Neyman orthogonality of the DR expressions \eqref{DiDidentDRcont2modified} and \eqref{DiDidentDRcont3modified} for panel data.
Consider the identification of the ATET $\Delta_{d_t,d_t', t}=E[Y_t(d_t)-Y_t(d_t')|D_t=d_t]$ in equation \eqref{ATEsameperiod} based on Assumptions \ref{ass4} to \ref{ass7}. We define the nuisance parameters $$\eta=(m_{d_t'}(t,\mathbf{D}_{t-1};X),p_{d_t}(X))$$ and denote the true values of the nuisance parameters in the population by $$\eta^0=(m^0_{d_t'}(t,\mathbf{D}_{t-1},X),p^0_{d_t}(\mathbf{D}_{t-1},X)).$$ We define $\Psi_{d'}=E[m_{d_t'}(t,\mathbf{D}_{t-1},X)|D_t=d_t]$ to be the average of the conditional mean difference $m_{d_t'}(t,\mathbf{D}_{t-1},X)=E[Y_{t}-Y_{t-1}| D_t=d_t', \mathbf{D}_{t-1},X]$ among those with treatment dose $D_t=d_t$. The ATET provided in equation \eqref{DiDidentpostreg2modified} may be expressed as a function of $\Psi_{d'}$: 
\begin{align}\label{nuisanceparmod}
&\Delta_{d_t,d_t', t}=E[  m_{d_t}(t,\mathbf{D}_{t-1},X) - m_{d_t'}(t,\mathbf{D}_{t-1},X)|D_t=d_t]\notag\\
&=E[Y_t-Y_{t-1}|D_t=d_t]-\Psi_{d_t'},
\end{align}
where the second equality follows from the law of iterated expectations. Therefore, considering a Neyman-orthogonal function for $\Psi_{d_t'}$ yields a Neyman-orthogonal expression of the ATET, which is an additive function of $\Psi_{d_t'}$. 

We denote by $\phi_{d_t'}$ the Neyman-orthogonal score function for $\Psi_{d_t'}$, which corresponds to the following expression:    
\begin{align}\label{neymanscore2}
\phi_{d_t'}&= \frac{  \omega(D_t;d_t, h) \cdot m_{d_t'}(t,\mathbf{D}_{t-1},X)}{P_{d_t}}\notag\\
&+\frac{\omega(D_t;d_t', h) \cdot p_{d_t}(\mathbf{D}_{t-1}, X)\cdot [Y_t-Y_{t-1}- m_{d_t'}(t,\mathbf{D}_{t-1},X)]}{ p_{d'_t}(\mathbf{D}_{t-1}, X)\cdot P_{d_t}}-\Psi_{d_t'}.
\end{align}

The Gateaux derivative of $E[\phi_{d_t'}]$ in the direction of $[\eta^0-\eta]$ is given by:
\begin{align}\label{derivscore2}
&\partial E[\phi_{d',t'}] [\eta-\eta^0] = \lim_{h \rightarrow 0} E \left[  \frac{  \omega(D_t;d_t, h) \cdot [m_{d_t'}(t,\mathbf{D}_{t-1},X)-m^0_{d_t'}(t,\mathbf{D}_{t-1},X)]}{ P_{d_t}}\right]\notag\\
&-\lim_{h \rightarrow 0} E \left[  \frac{  \omega(D_t;d_t', h) \cdot p^0_{d_t}(\mathbf{D}_{t-1},X) \cdot [m_{d_t'}(t,\mathbf{D}_{t-1},X)-m^0_{d_t'}(t,\mathbf{D}_{t-1},X)]}{ p^0_{d'_t}(\mathbf{D}_{t-1}, X) \cdot P_{d_t}}\right] \\
&+\lim_{h \rightarrow 0} E \left[  \frac{  \omega(D_t;d_t', h) \cdot [Y_t-Y_{t-1}- m^0_{d_t'}(t,\mathbf{D}_{t-1},X)]}{ p^0_{d'_t}(\mathbf{D}_{t-1}, X) } \cdot \frac{[ p_{d_t}(\mathbf{D}_{t-1}, X) - p^0_{d_t}(\mathbf{D}_{t-1}, X)] } {P_{d_t}}\right]\notag\\
&-\lim_{h \rightarrow 0} E \left[  \frac{  \omega(D_t;d_t', h) \cdot p^0_{d_t}(\mathbf{D}_{t-1}, X)\cdot [Y_t-Y_{t-1}- m^0_{d_t'}(t,\mathbf{D}_{t-1},X)]}{ p^0_{d'_t}(\mathbf{D}_{t-1}, X) \cdot P_{d_t} } \cdot \frac{ p_{d'_t}(\mathbf{D}_{t-1},X) -  p^0_{d'_t}(\mathbf{D}_{t-1},X)} { p^0_{d'_t}(\mathbf{D}_{t-1},X) }\right]=0.\notag
\end{align}

Analogously to expression \eqref{derivscore}, applying the law of iterated expectations to the term in the second line of equation \eqref{derivscore2} yields:
\begin{align}
&-\lim_{h \rightarrow 0} E \left[  \frac{  E[\omega(D_t;d_t', h)|\mathbf{D}_{t-1},X] \cdot p^0_{d_t}(\mathbf{D}_{t-1},X) \cdot [m_{d_t'}(t,\mathbf{D}_{t-1},X)-m^0_{d_t'}(t,\mathbf{D}_{t-1},X)]}{ p^0_{d'_t}(\mathbf{D}_{t-1}, X) \cdot P_{d_t}}\right]\notag\\
&=- E \left[  \frac{   p^0_{d_t}(\mathbf{D}_{t-1},X) \cdot [m_{d_t'}(t,\mathbf{D}_{t-1},X)-m^0_{d_t'}(t,\mathbf{D}_{t-1},X)]}{ P_{d_t}}\right].\notag
\end{align}
Applying the law of iterated expectations to the term on the left hand side of the first line yields:
\begin{align}
 &\lim_{h \rightarrow 0} E \left[  \frac{  E[\omega(D_t;d_t, h)|\mathbf{D}_{t-1},X] \cdot [m_{d_t'}(t,\mathbf{D}_{t-1},X)-m^0_{d_t'}(t,\mathbf{D}_{t-1},X)]}{ P_{d_t}}\right]\notag\\
 & = E \left[  \frac{   p^0_{d_t}(\mathbf{D}_{t-1},X) \cdot [m_{d_t'}(t,\mathbf{D}_{t-1},X)-m^0_{d_t'}(t,\mathbf{D}_{t-1},X)]}{ P_{d_t}}\right].\notag
\end{align}
Consequently, the terms in the first and in the second line of equation \eqref{derivscore2} cancel out. Considering the third line, we have that:
\begin{align}
&\lim_{h \rightarrow 0} E \left[  E \left[  \frac{  \omega(D_t;d_t', h) \cdot [Y_t-Y_{t-1}- m^0_{d_t'}(t,\mathbf{D}_{t-1},X)]}{ p^0_{d'_t}(\mathbf{D}_{t-1}, X) } \cdot \frac{[ p_{d_t}(\mathbf{D}_{t-1}, X) - p^0_{d_t}(\mathbf{D}_{t-1}, X)] } {P_{d_t}}\bigg| \mathbf{D}_{t-1},X\right] \right] \notag\\
&= [m^0_{d_t'}(\mathbf{D}_{t-1}, X)-m^0_{d_t'}(\mathbf{D}_{t-1}, X)] \cdot \frac{[ p_{d_t}(\mathbf{D}_{t-1}, X) - p^0_{d_t}(\mathbf{D}_{t-1}, X)] } {P_{d_t}} =0. \notag
\end{align}
Concerning the fourth line, 
\begin{align}
-& \lim_{h \rightarrow 0} E \left[  E \left[  \frac{  \omega(D_t;d_t', h) \cdot p^0_{d_t}(\mathbf{D}_{t-1}, X)\cdot [Y_t-Y_{t-1}- m^0_{d_t'}(t,\mathbf{D}_{t-1},X)]}{ p^0_{d'_t}(\mathbf{D}_{t-1}, X) \cdot P_{d_t} }\right.\right.\notag\\ &\times \left.\left. \frac{ p_{d'_t}(\mathbf{D}_{t-1},X) -  p^0_{d'_t}(\mathbf{D}_{t-1},X)} { p^0_{d'_t}(\mathbf{D}_{t-1},X) }\bigg| \mathbf{D}_{t-1}, X\right]\right]\notag\\
=&- E \left[  \frac{  p^0_{d_t}(\mathbf{D}_{t-1}, X) \cdot [m^0_{d_t'}(\mathbf{D}_{t-1}, X)-m^0_{d_t'}(\mathbf{D}_{t-1}, X)]}{   P_{d_t} } \cdot \frac{ p_{d'_t}(\mathbf{D}_{t-1},X) -  p^0_{d'_t}(\mathbf{D}_{t-1},X)} { p^0_{d'_t}(\mathbf{D}_{t-1},X) } \right]=0.\notag
\end{align}
Thus, the Gateaux derivative of $E[\phi_{d'_t}]$ in the direction of $[\eta^0-\eta]$ is zero. Furthermore, the Gateaux derivative of $E[Y_t-Y_{t-1}|D_t=d_t]$ in the direction of $[\eta^0-\eta]$ is zero. We note that the DR expression \eqref{DiDidentDRcont2modified} is equivalent to expression \eqref{nuisancepar} while using the Neyman-orthogonal score functions \eqref{neymanscore2} for computing $\Psi_{d'_t}$, noting that $E[Y_t-Y_{t-1}|D_t=d_t]=\lim_{h \rightarrow 0} E \left[\frac{  \omega(D;d, h) \cdot (Y_t-Y_{t-1})}{P_{d_t}}\right]$. As a result, the DR expression \eqref{DiDidentDRcont2modified} satisfies the Neyman orthogonality.

Lastly, consider the identification of the ATET $\Delta_{d_{t-s},d'_{t-s}, t}=E[Y_t(d_{t-s})-Y_t(d'_{t-s})|D_{t-s}=d_{t-s}]$ in equation \eqref{ATElagged} based on Assumptions \ref{ass4}, \ref{ass6}, \ref{ass7}, and \ref{ass8} using DR expression \eqref{DiDidentDRcont3modified} for panel data. In this case, the steps for showing Neyman orthogonality are analogous. The only difference to the previous is the introduction of a time lag between the treatment and post-treatment outcomes.

\subsection{Proofs of asymptotic theories}\label{asymproof}
In this section, we provide the main ideas of the proofs for our asymptotic results and omit the details, which closely follow \cite{Chetal2018}, \cite{Chang2020}, and \cite{zhang2025}. We focus on the repeated cross-sectional case and note that the arguments for panel case follows from \cite{zhang2025} with minor notation changes.

First, we outline the main idea for Lemma \ref{lemma:bias}. Recall that when introducing the regularity conditions in Section \ref{asymp}, we have defined the bias as the difference between the true ATET and its smoothed kernel counterpart when the continuous treatment intensity is approximated by a kernel function with bandwidth $h$:
\begin{align*}
    B(h) = \Delta_{d_t,d_t',t} - \Delta_{d_t,d_t',t}^h.
\end{align*}
In the repeated cross-sectional case, this bias can be shown to depend on the difference between the conditional density and its kernel approximation:
\begin{align*}
    \rho_{d_t}(D_{t-1}, X) - \rho_{d_t}^h( D_{t-1}, X).
\end{align*}
In particular, it can be shown that
\begin{align*}
    \rho_{d_t}^h( D_{t-1}, X) & = E[\omega(D_t;d_t,h)\cdot I\{T=t\}|\mathbf{D}_{t-1},X]\\
    & = P(T=t|\mathbf{D}_{t-1},X)\int \frac{1}{h}\mathcal{K}\left(\frac{D_t-d_t}{h}\right) \rho_s(D_{t-1},T=t, X) ds\\
    & \asymp \rho_d(D_{t-1},t, X) + Ch^2 \rho_d^{(2)}(D_{t-1},t, X) + o(h^2).
\end{align*}
Therefore, taking the difference $\rho_d(D_{t-1},t,X) - \rho_d^h( D_{t-1},t,X)$, by the assumptions on the kernel function and the conditional density, the bias is of order $O(h^2)$.

Second, to establish asymptotic normality in Theorem \ref{thm:normality}, we consider the following decomposition:
\begin{align*}
    \hat{\Delta}_{d_t,d_t',t} - \Delta_{d_t,d_t',t} &= \hat{\Delta}_{d_t,d_t',t} - \Delta_{d_t,d_t',t}^h + \Delta_{d_t,d_t',t} - \Delta_{d_t,d_t',t}^h\\
    &= \hat{\Delta}_{d_t,d_t',t} - \Delta_{d_t,d_t',t}^h + B(h).
\end{align*}
It suffices to focus on the difference between our ATET estimator and the kernel counterpart of the population ATET. In particular, the Neyman orthogonality results By the definition of the score function \eqref{scorerepeated} in Section \ref{sec:meth} and using Taylor's theorem, we have the following expansion:
\begin{align}
    \sqrt{n}(\hat{\Delta}_{d_t,d_t',t}^h - \Delta_{d_t,d_t',t}^h) & = \sqrt{n} \frac{1}{K}\sum_{k=1}^K E_{nk}[\psi^h(W,\Delta_{d_t,d_t',t}^h, \Pi_{d_t,t}, \hat{\eta}_k) ] \label{eq:1}\\
    &+ \sqrt{n} \frac{1}{K}\sum_{k=1}^K E_{nk} [\partial_\Pi \psi^h(W,\Delta_{d_t,d_t',t}^h, \Pi_{d_t,t}, \hat{\eta}_k) ] (\hat{\Pi}_{d_t,t}^k - \Pi_{d_t,t})\label{eq:2}\\
    &+ \sqrt{n}\frac{1}{K}\sum_{k=1}^K E_{nk}[\partial_\Pi^2 \psi^h(W,\Delta_{d_t,d_t',t}^h, \bar{\Pi}, \hat{\eta}_k) ] (\hat{\Pi}_{d_t,t}^k - \Pi_{d_t,t})^2 \label{eq:3}
\end{align}
where the derivative $\partial_\Pi$ is w.r.t. the density $\Pi_{d_t,t}$, and $\bar{\Pi} \in (\hat{\Pi}_{d_t,t}^k, \Pi_{d_t,t})$. We use $E_{nk}$ to denote the empirical average with subsample $\mathcal{W}_k$. Moreover, due to cross-fitting, we can treat the high-dimensional nuisance parameters $\hat{\eta}_k$ as fixed without loss of generality. This significantly lessens the requirement on the complexity of the nuisance parameter space and greatly simplifies the asymptotic analysis. Then, the asymptotic result follows from standard arguments and we only provide a high-level summary.  that by the standard arguments on kernel density, we have $|\hat{\Pi}_{d_t,t}^k - \Pi_{d_t,t}| = O_p((nh)^{-1/2})$ and $|\hat{\Pi}_{d_t,t}^k - \Pi_{d_t,t}|^2 = O_p((nh)^{-1})$. 

In the first step, we can show that the second-order term \eqref{eq:3} vanishes asymptotically and does not contribute to the asymptotic variance. Specifically, by the standard results for kernel density estimators, $|\hat{\Pi}_{d_t,t}^k - \Pi_{d_t,t}|^2 = O_p((nh)^{-1})$. Moreover, it can be shown that
\begin{align}
|E_{nk}[\partial_\Pi^2 \psi^h(W,\Delta_{d_t,d_t',t}^h, \bar{\Pi}, \hat{\eta}_k) ] - E[\partial_\Pi^2 \psi^h(W,\Delta_{d_t,d_t',t}^h, \Pi_{d_t,t}, \eta_0) ]| = O_p((nh)^{-1/2} + h^{-1/2}\varepsilon_n).
\end{align}
Then by the boundness and rate assumptions, we conclude that \eqref{eq:3} is $o_p(1)$.

In the second step, we bound the first-order term \eqref{eq:2} and show that this term contributes to the asymptotic variance. It can be shown that
\begin{align}
 E_{nk} [\partial_\Pi \psi^h(W,\Delta_{d_t,d_t',t}^h, \Pi_{d_t,t}, \hat{\eta}_k)] = E[\partial_\Pi \psi^h(W,\Delta_{d_t,d_t',t}^h, \Pi_{d_t,t}, \eta_0) ] + o_p(1)
\end{align}
if $h^{-1}\varepsilon_n^2 = o(1)$. Then using the facts that $|\hat{\Pi}_{d_t,t}^k - \Pi_{d_t,t}| = O_p((nh)^{-1/2})$ and 
\begin{align}
    \hat{\Pi}_{d_t,t}^k - \Pi_{d_t,t} = \frac{1}{n-n_k}\sum_{i\in\mathcal{W}_k^c} \omega(D_{i,t}; d_t, h)\cdot I\{T_i =t\} - E[\omega(D_{i,t}; d_t, h)\cdot I\{T_i =t\} + O(h^2),
\end{align}
we conclude that 
\begin{align}
    \eqref{eq:2}= \frac{1}{n}\sum_{i=1}^n \frac{\Delta_{d_t,d_t',t}^h}{\Pi_{d_t,t}}( \omega(D_{i,t}; d_t, h)\cdot I\{T_i =t\} - E[\omega(D_{i,t}; d_t, h)\cdot I\{T_i =t\}) + o_p((Nh)^{-1/2})
\end{align}
where we have used the fact that $E[\partial_\Pi \psi^h(W,\Delta_{d_t,d_t',t}^h, \Pi_{d_t,t}, \eta_0) ] = \Delta_{d_t,d_t',t}^h/\Pi_{d_t,t}$.

In the third step, we expand \eqref{eq:1} around the nuisance estimator $\hat{\eta}$ and use Neyman orthogonality to show that the first-order bias disappears and second-order terms are negligible. Specifically, \eqref{eq:1} can be decomposed as the following
\begin{align}
    \eqref{eq:1} = \frac{1}{\sqrt{n}}\sum_{i=1}^n \psi^h(W_i,\Delta_{d_t,d_t',t}^h, \Pi_{d_t,t}, \eta_0) + \sqrt{n}\frac{1}{K}\sum_{k=1}^K R_{nk}
\end{align}
where 
\begin{align}
    R_{nk} = E_{nk}[\psi^h(W,\Delta_{d_t,d_t',t}^h, \Pi_{d_t,t}, \hat{\eta}_k)] - E_{nk}[\psi^h(W,\Delta_{d_t,d_t',t}^h, \Pi_{d_t,t}, \eta_0)].
\end{align}
To bound $R_{nk}$, we can apply triangle inequality, with $n_k := n/K$ denoting the size of the subsample,
\begin{align*}
    |R_{nk}| \leq (R_{1k} + R_{2k})/\sqrt{n_k}
\end{align*}
where
\begin{align}
    R_{1k} &= | G_{nk}[\psi^h(W,\Delta_{d_t,d_t',t}^h, \Pi_{d_t,t}, \hat{\eta}_k)] - G_{nk}[\psi^h(W,\Delta_{d_t,d_t',t}^h, \Pi_{d_t,t}, \eta_0)]|\\
    R_{2k} &= \sqrt{n_k}|E[\psi^h(W,\Delta_{d_t,d_t',t}^h, \Pi_{d_t,t}, \hat{\eta}_k)|\mathcal{W}_k^c] - E[\psi^h(W,\Delta_{d_t,d_t',t}^h, \Pi_{d_t,t}, \eta_0)]|.
\end{align}
Specifically, it can be shown that $R_{1k} = O_p(h^{-1}\varepsilon_n)$ by sample splitting, the assumptions on the nuisance estimators, the law of iterated expectation, and conditional Markov's inequality. Moreover, by Neyman orthogonality, it can be shown that $R_{2k} = O(n^{1/2}h^{-1/2}\varepsilon_n^2)$. Therefore, 
\begin{align}
        \eqref{eq:1} = \frac{1}{\sqrt{n}}\sum_{i=1}^n \psi^h(W_i,\Delta_{d_t,d_t',t}^h, \Pi_{d_t,t}, \eta_0) + o_p(h^{-1/2}).
\end{align}

In the final step, combining above results, the asymptotic normality follows from a triangular array central limit theorem.

Lastly, to establish the consistency of our variance estimator as in Theorem \ref{thm:var}, we define:
\begin{align*}
    \tilde{\phi}(W, \Delta_{d_t,d_t',t}^h, \Pi_{d_t,t}, \eta) =  \phi^h(W,\Delta_{d_t,d_t',t}^h, P_{d_t},\eta) -  \frac{\Delta_{d_t, d_t', t}^h}{\Pi_{d_t,t}}\left(\omega(D_t;d_t,h) - E[\omega(D_t;d_t,h)]\right).
\end{align*}
Note that by definition,
\begin{align*}
    \sigma_h^2  = E[ \tilde{\phi}^2(W, \Delta_{d_t,d_t',t}^h, \Pi_{d_t,t},\eta) ];\quad \hat{\sigma}_h^2 = E_n[ \tilde{\phi}^2(W, \hat{\Delta}_{d_t,d_t',t}, \hat{\Pi}_{d_t,t}, \hat{\eta})].
\end{align*}
It suffices to show that
\begin{align*}
    J_n := | E_n[ \tilde{\phi}^2(W, \hat{\Delta}_{d_t,d_t',t}, \hat{\Pi}_{d_t,t}, \hat{\eta})] - E[ \tilde{\phi}^2(W, \Delta_{d_t,d_t',t}^h, \Pi_{d_t,t},\eta) ]| = o_p(1).
\end{align*}
Using the triangle inequality, we have:
\begin{align}
    J_n &\leq | E_n[ \tilde{\phi}^2(W, \hat{\Delta}_{d_t,d_t',t}, \hat{\Pi}_{d_t,t}, \hat{\eta})] - E_n[ \tilde{\phi}^2(W, \Delta_{d_t,d_t',t}^h, \Pi_{d_t,t},\eta) ]|\label{eq:4}\\
    & + | E_n[ \tilde{\phi}^2(W, \Delta_{d_t,d_t',t}^h, \Pi_{d_t,t},\eta) ] - E[ \tilde{\phi}^2(W, \Delta_{d_t,d_t',t}^h, \Pi_{d_t,t},\eta) ]|.\label{eq:5}
\end{align}
First, we can show that \eqref{eq:5} is $o_p(1)$ using the assumptions on the kernel function and boundedness. Moreover, the term \eqref{eq:4}, which we denote by $J_{1n}$, satisfies:
\begin{align*}
    J_{1n}^2 \lesssim S_n\left(S_n + \frac{1}{n}\sum_{i=1}^n \tilde{\phi}^2(W_i, \Delta_{d_t,d_t',t}^h, \Pi_{d_t,t},\eta)\right),
\end{align*}
where 
\begin{align*}
    S_n = \frac{1}{n}\sum_{i=1}^n |\tilde{\phi}(W_i, \Delta_{d_t,d_t',t}^h, \Pi_{d_t,t},\eta) -  \tilde{\phi}(W, \hat{\Delta}_{d_t,d_t',t}, \hat{\Pi}_{d_t,t}, \hat{\eta})|^2.
\end{align*}
Specifically, $S_n$ can be shown to be $O_p(h^{-1}\varepsilon_n^2 + h^{-2}n^{-1})$, which follows from an expansion of the term $\tilde{\phi}(W, \hat{\Delta}_{d_t,d_t',t}, \hat{\Pi}_{d_t,t}, \hat{\eta})$ around $\hat{\Delta}_{d_t,d_t',t}$ using Taylor's theorem. Moreover, the remaining term in $J_{1n}^2$, namely $n^{-1}\sum_{i=1}^n \tilde{\phi}^2(W_i, \Delta_{d_t,d_t',t}^h, \Pi_{d_t,t},\eta)$, is of order $O_p(h^{-1})$ by our assumptions on the kernel function and boundedness. Therefore, $J_{1n}$ is $o_p(1)$ if $h^{-2}\varepsilon_n^2 + h^{-3}n^{-1} = o(1)$, and the consistency of our variance estimator follows. $\blacksquare$

\subsection{Descriptive statistics}\label{stats_did}

\begin{footnotesize}
\setlength{\tabcolsep}{3pt} 
\begin{longtable}{l ccccccc}
    \caption{
    Descriptive statistics for  extended set of variables} \label{tab:full_appendix} \\
    \hline\hline
    variable & mean& median& std. dev.& skewness & kurtosis & minimum & maximum\\
    \hline
    \endfirsthead
    
    \multicolumn{8}{c}{{ \tablename\ \thetable{} -- continued from previous page}} \\
    \hline
    variable & mean& median& std. dev.& skewness & kurtosis & minimum & maximum\\
    \hline
    \endhead

    \hline
    \multicolumn{8}{r}{{continues on next page...}} \\
        \endfoot

    \bottomrule
    \endlastfoot

    \multicolumn{8}{l}{\textit{Panel A: COVID-19 outcomes}} \\
    deaths / population & 0.0020 & 0.0019 & 0.0014 & 0.50 & -0.62 & 0.0000 & 0.0075 \\
    \addlinespace
    
    \multicolumn{8}{l}{\textit{Panel B: vaccination}} \\
    dose-1 vaccination & 0.26 & 0.11 & 0.30 & 0.79 & -0.98 & 0.00 & 1.30 \\
    dose-2 vaccination & 0.36 & 0.24 & 0.35 & 0.27 & -1.65 & 0.00 & 1.07 \\
    \addlinespace

    \multicolumn{8}{l}{\textit{Panel C: Human Development Index (MHDI)}} \\
    municipal HDI (overall) & 0.74 & 0.75 & 0.05 & -0.69 & 0.77 & 0.57 & 0.86 \\
    longevity dimension & 0.84 & 0.84 & 0.03 & -0.74 & 0.62 & 0.72 & 0.89 \\
    education dimension & 0.67 & 0.68 & 0.07 & -0.61 & 0.45 & 0.42 & 0.81 \\
    income dimension & 0.73 & 0.74 & 0.06 & -0.34 & 0.73 & 0.56 & 0.89 \\
    \addlinespace

    \multicolumn{8}{l}{\textit{Panel D: inequality \& poverty}} \\
    Gini index & 0.50 & 0.50 & 0.05 & -0.09 & -0.15 & 0.35 & 0.63 \\
    Theil-L index & 0.46 & 0.45 & 0.10 & 0.34 & -0.01 & 0.21 & 0.76 \\
    proportion of poor & 8.69 & 5.48 & 8.08 & 1.83 & 3.25 & 0.44 & 43.46 \\
    proportion extremely poor & 2.72 & 1.29 & 3.67 & 2.82 & 10.08 & 0.00 & 25.72 \\
    vulnerable to poverty (\%) & 24.00 & 19.88 & 14.17 & 1.10 & 0.60 & 3.85 & 69.10 \\
    ratio 10\% rich / 40\% poor & 13.58 & 12.79 & 4.10 & 0.92 & 1.03 & 5.98 & 28.23 \\
    per capita income & 807.73 & 792.54 & 282.31 & 1.12 & 2.75 & 257.64 & 2043.74 \\
    income approp. (Top 10\%) & 40.75 & 40.70 & 4.86 & 0.06 & -0.23 & 28.77 & 53.15 \\
    income approp. (Bottom 40\%) & 12.65 & 12.61 & 2.23 & 0.14 & -0.08 & 6.98 & 19.26 \\
    \addlinespace

    \multicolumn{8}{l}{\textit{Panel E: education \& literacy (\%)}} \\
    complete secondary (25+) & 36.20 & 36.22 & 8.89 & 0.14 & 0.17 & 15.59 & 64.08 \\
    complete primary (25+) & 51.78 & 52.44 & 9.90 & -0.30 & 0.00 & 23.79 & 76.83 \\
    complete higher ed (25+) & 11.16 & 10.59 & 5.05 & 0.99 & 1.91 & 1.57 & 31.86 \\
    illiteracy rate (11-14) & 1.90 & 1.34 & 1.68 & 3.19 & 13.20 & 0.00 & 13.04 \\
    illiteracy rate (15+) & 7.26 & 5.65 & 5.07 & 2.15 & 5.14 & 1.55 & 29.69 \\
    \addlinespace

    \multicolumn{8}{l}{\textit{Panel F: school attendance rates (\%)}} \\
    age 0 to 3 & 27.66 & 26.28 & 10.20 & 0.45 & -0.17 & 7.31 & 56.87 \\
    age 15 to 17 & 84.36 & 85.13 & 4.35 & -0.56 & 0.13 & 72.25 & 95.56 \\
    age 18 to 24 & 29.68 & 30.16 & 5.94 & -0.09 & -0.10 & 12.81 & 45.85 \\
    age 4 to 5 & 83.40 & 85.02 & 10.34 & -0.96 & 0.76 & 43.36 & 98.88 \\
    age 6 to 14 & 97.51 & 97.61 & 1.09 & -0.93 & 2.15 & 92.37 & 99.76 \\
    net enrollment primary & 92.57 & 92.61 & 1.77 & -0.10 & 0.46 & 86.52 & 97.76 \\
    net enrollment secondary & 49.03 & 49.53 & 9.31 & -0.18 & -0.48 & 23.94 & 72.90 \\
    net enrollment higher ed & 15.85 & 15.53 & 6.42 & 0.37 & 0.07 & 2.74 & 35.89 \\
    \addlinespace

    \multicolumn{8}{l}{\textit{Panel G: demographics \& survival}} \\
    total population 2022 & {211,866} & {103,440} & {306,890} & 4.11 & 25.44 & {2,892} & {2,817,381} \\
    urban population (prop.) & 0.90 & 0.93 & 0.10 & -1.94 & 5.18 & 0.36 & 1.00 \\
    rural Population (prop.) & 0.10 & 0.07 & 0.10 & 1.94 & 5.18 & 0.00 & 0.64 \\
    aging rate & 11.46 & 11.67 & 2.71 & -0.22 & -0.11 & 3.67 & 18.73 \\
    total fertility rate & 1.81 & 1.79 & 0.30 & 1.07 & 2.46 & 1.25 & 3.07 \\
    dependency ratio & 43.71 & 43.30 & 2.95 & 0.38 & -0.33 & 37.64 & 52.13 \\
    prob. survival to 40 yrs & 94.44 & 94.64 & 1.18 & -0.68 & 0.04 & 91.11 & 96.66 \\
    prob. survival to 60 yrs & 84.56 & 84.83 & 2.22 & -0.36 & 0.13 & 77.37 & 89.79 \\
    life expectancy at birth & 75.20 & 75.42 & 1.72 & -0.74 & 0.62 & 68.36 & 78.44 \\
    under-five mortality rate & 16.48 & 15.63 & 3.73 & 1.78 & 5.96 & 9.98 & 38.14 \\
    infant mortality rate & 14.55 & 13.80 & 3.59 & 1.80 & 5.66 & 8.49 & 35.00 \\
    \addlinespace

    \multicolumn{8}{l}{\textit{Panel H: housing \& infrastructure (\%)}} \\
    households piped water & 95.63 & 97.80 & 5.56 & -2.98 & 12.26 & 60.18 & 99.99 \\
    households density $>$ 2 & 22.82 & 22.00 & 9.03 & 0.62 & 0.06 & 6.55 & 51.40 \\
    households electricity & 99.72 & 99.90 & 0.50 & -3.75 & 16.50 & 96.48 & 100.00 \\
    households garbage coll.& 98.35 & 99.34 & 2.72 & -3.59 & 17.34 & 78.10 & 100.00 \\
    households sewage & 68.49 & 78.72 & 28.91 & -0.93 & -0.34 & 0.71 & 99.83 \\
    sewage / septic tank & 84.60 & 90.51 & 17.77 & -2.05 & 4.28 & 5.30 & 99.95 \\
    inadequate water/sewage & 2.55 & 0.63 & 4.83 & 3.17 & 10.72 & 0.00 & 27.16 \\
    non-masonry walls & 1.54 & 0.65 & 2.76 & 4.87 & 32.22 & 0.00 & 25.84 \\
\end{longtable}
\end{footnotesize}

\end{document}